\documentclass[a4paper,12pt]{article}
\usepackage[latin1]{inputenc}
\usepackage{graphicx}
\usepackage[font=small]{caption}
\usepackage{amsmath,amsfonts,amssymb,amscd}
\usepackage[usenames,dvipsnames]{pstricks}
\usepackage[refpage]{nomencl} 
\usepackage{pst-grad} 
\usepackage{pst-plot} 
\usepackage{theorem}
\usepackage{enumerate,epsfig}
\usepackage[all,cmtip]{xy} 
\usepackage{float}

\usepackage{tabularx}
\newcolumntype{C}[1]{>{\centering\arraybackslash}p{#1}} 
\newcolumntype{R}[1]{>{\raggedleft\arraybackslash}p{#1}} 

\usepackage{pstricks, pst-node, pst-tree}
\usepackage{hyperref} 
\hypersetup{ 
 pdfpagemode=FullScreen,  
    unicode=false,          
    pdftoolbar=true,        
    pdfmenubar=true,        
    pdffitwindow=false,     
    pdfstartview={FitH},    
    pdftitle={Properties of Persistent Mutual Information and Emergence},    	
    pdfauthor={Peter Gmeiner},     	
    pdfsubject={Subject},   
    pdfcreator={Peter Gmeiner},   	
    pdfproducer={Producer}, 
    pdfkeywords={Excess entropy, persistent mutual information, emergence}, 
    pdfnewwindow=true,      
    colorlinks=true,        
    linkcolor=black,        
    citecolor=blue,        
    filecolor=black,        
    urlcolor=black          
}

\usepackage{fancyhdr} 

\theoremstyle{change}
\newtheorem{Lemma}{Lemma}[section]

\newtheorem{Def}[Lemma]{Definition}
\newtheorem{Satz}[Lemma]{Proposition}

\newtheorem{Korollar}[Lemma]{Corollary}
{\theorembodyfont{\normalfont}\newtheorem{Bsp}[Lemma]{Example}}
{\theorembodyfont{\normalfont}}
{\theorembodyfont{\normalfont}\newtheorem{Bem}[Lemma]{Remark}}
\theoremstyle{changebreak}

\newcommand{\setN}{\mathbb{N}} 
\newcommand{\setZ}{\mathbb{Z}} 
\newcommand{\setR}{\mathbb{R}} 
\newcommand{\setC}{\mathbb{C}} 
 
\newcommand{\setId}{\mathbb{I}} 
\newcommand{\kA}{\mathcal{A}}	
\newcommand{\kC}{\mathcal{C}}	
\newcommand{\kF}{\mathcal{F}}	
\newcommand{\kG}{\mathcal{G}}	
\newcommand{\kM}{\mathcal{M}}	
\newcommand{\kS}{\mathcal{S}}	
\newenvironment{Proof}{ \par {\it Proof.} \par }{\hspace*{\fill}
$\Box$\par\vskip2ex}	

\def\XXint#1#2#3{{\setbox0=\hbox{$#1{#2#3}{\int}$}
     \vcenter{\hbox{$#2#3$}}\kern-.5\wd0}}

\newcommand{\nt}[1]{{\bf #1}\index{#1}}

\begin{document}

\numberwithin{equation}{section}

\title{Properties of Persistent Mutual Information and Emergence}
\author{Peter Gmeiner\footnote{Department Mathematik,
Friedrich-Alexander-Universit\"at Erlangen-N\"urnberg, Cauerstr. 11,
91058 Erlangen, Germany. E-mail: gmeiner@mi.uni-erlangen.de}}
\date{\today}
\maketitle
\begin{abstract}
The persistent mutual information (PMI) is a complexity measure for stochastic 
processes. It is related to well-known complexity measures like excess entropy
or statistical complexity. Essentially it is a variation of the excess entropy
so that it can be interpreted as a specific measure of system internal memory.
The PMI was first introduced in 2010 by Ball, Diakonova and MacKay as a measure
for (strong) emergence \cite{Bal10}. In this paper we define the PMI
mathematically and investigate the relation to excess entropy and statistical
complexity. In particular we prove that the excess entropy is an upper bound of
the PMI. Furthermore we show some properties of the PMI and calculate it
explicitly for some example processes. We also discuss to what extend it is a
measure for emergence and compare it with alternative approaches used to
formalize emergence.
\end{abstract}

\section{Preliminaries}

Let $(\Omega, \kF, P)$ be a probability space with a metric space $\Omega$, a
$\sigma$-algebra $\kF$ and a probability measure $P$. 
For random variables $X, Y: \Omega \rightarrow \kA$ mapping to a finite
alphabet $\kA$ the Shannon entropy is defined by
$$H(X) := - \sum_{x \in \kA} \Pr(X = x) \log \Pr(X = x),$$
and the conditioned Shannon entropy by
$$H(X \mid Y) := - \sum_{x, y \in \kA} \Pr(X = x, Y = y) \log \Pr(X = x \mid Y =
y),$$
where $\Pr(X=x) := P\left ( \left \{\omega \in \Omega \mid X(\omega) = x \right
\}
\right )$ denotes the probability that the random variable $X$ is equal to $x
\in \kA$, $\Pr(X=x, Y=y)$ is the joint probability between $X$ and $Y$ and for
$\Pr(Y=y) > 0$ the conditional probability is $\Pr(X=x\mid Y=y) :=
\frac{\Pr(X=x,
Y=y)}{\Pr(Y=y)}$.
In the definitions the convention $0 \log(0) = 0$ is used. The mutual
information between two random variables is 
$$I(X;Y) := H(X) - H(X\mid Y).$$
The mutual information is non-negative $(I(X;Y) \ge 0)$ and equals zero if and
only if $X$ and $Y$ are independent random variables \cite{Cov06}.

We consider a time-discrete stationary stochastic process
$\overleftrightarrow{S} := (S_t)_{t \in \setZ}$ with random variables $S_t :
\Omega \rightarrow \kA$ for times $t \in \setZ$. We define the semi-infinite
processes $\overleftarrow{S} := (S_{-t})_{t \in \setN}$ interpreted as past and
$\overrightarrow{S} := (S_t)_{t \in \setN_0}$ interpreted as future
respectively. Blocks of random variables with finite length are denoted by
$S_a^b := (S_k)_{k\in [a,b]\cap \setZ}$ for $-\infty < a \le b < \infty$ and
the corresponding block entropy is $H(L) := H(S_1^L) =H(S_1, \ldots, S_L)$. The
one-sided sequence space is $\kA^\setN := \times_{i \in \setN} \kA$ and in
the same way the two-sided sequence space $\kA^\setZ$ is defined. We
introduce the shift function $\sigma: \kA^\setZ \rightarrow \kA^\setZ$ by
$\sigma(x)_i := x_{i+1}$.
At any time $t \in \setZ$ we have random variables $S_{-\infty}^t := (S_k)_{k
\le t}$ and $S_{t+1}^\infty:=(S_k)_{k \ge t+1}$ that govern the systems observed
behaviour respectively in the shifted past and the shifted future. The mutual
information between these two variables is the well-known {\em excess entropy}
\cite{Cru83, Cru03}
\begin{equation} \label{ExcessDef}
E := \lim_{L \rightarrow \infty} I(S_0^{L-1} ; S_{-L}^{-1}).
\end{equation}
In general, it is not clear if the limit in (\ref{ExcessDef}) exists (for
Markov processes of finite order one can prove the existence). With the
assumption that the limit in (\ref{ExcessDef}) exists as a finite number the
following equality holds: $E = I(\overleftarrow{S}; \overrightarrow{S})$, see
Chapter 2.2 in \cite{Pin64}.

\section{Conceptualization}

The definition of the excess entropy (\ref{ExcessDef}) allows a concrete
information theoretic interpretation. In particular the excess entropy can be
seen as a specific measure of {\it system internal memory}. We will take
this as a basis to define a new term, first suggested in \cite{Bal10}, which
will capture the structural
behavior of a dynamical system on the whole time-domain. In particular it should
be possible to detect any existing {\em inherent structure} of the system which
will survive for all times. In order to achieve this goal we adapt the mutual
information-based representation of the excess entropy and introduce the
following expression
$$E_{t,\tau}^L := I(S_t^{t+L-1}; S_{-\tau-L+1}^{-\tau} ).$$
For $t= 0$ and $\tau =1$ this expression coincide with the finite-length excess 
entropy and we have
$$E = \lim_{L \rightarrow \infty} E_{0,1}^L.$$
For arbitrary $t$ and $\tau$-values we get a family of similar terms
$$E_{t,\tau} := \lim_{L \rightarrow \infty} E_{t,\tau}^L.$$
Every expression $E_{t,\tau}$ is the excess entropy with a time-gap of size 
$|t- \tau|$ between a random variable block of the past and the future.
For stationary processes we can write $E_{t,\tau}^L$ as
$$E_{t, \tau}^L = I(S_0^{L-1}; S_{-\tau-t-L+1}^{-\tau -t}) = I(S_0^{L-1}; S_{-k-L+1}^{-k}),$$
with $k:= \tau + t$. Instead of $E_{t,\tau}^L$ we often write $E_k^L$. 

To ensure that the double sequence $(E_k^L)_{L,k \in \setN}$ converges to
$\tilde E$ (written as $\displaystyle \lim_{k,L \rightarrow \infty} E_k^L$),
for every $\epsilon >0$ two numbers $m,n \in \setN$ need to exist so that for
all $k > m, L > n$ holds that $|E_k^L - \tilde E | < \epsilon$. A {\it simple
sequence} in the double sequence $E_k^L$ is defined with two subsequences
$L_i \stackrel{i \rightarrow \infty}{\rightarrow} \infty$
and $k_i \stackrel{i \rightarrow \infty}{\rightarrow} \infty$ by $\left
(E_{k_i}^{L_i} \right )_{i \in \setN}$. The double sequence $E_k^L$ converge to 
$\tilde E$, if and only if all simple sequences in $E_k^L$ converge to $\tilde
E$ \cite{Lon00}. In particular it holds that
$$
\lim_{k \rightarrow \infty} \lim_{L \rightarrow \infty}
E_k^L = \lim_{L \rightarrow \infty} \lim_{k \rightarrow \infty} E_k^L.
$$
The reverse direction of the last conclusion does not hold.

\begin{Def}
Let a stochastic process with values in a finite alphabet $\kA$ be given. The
\nt{persistent mutual information} of such a process is defined by
$$
PMI := \lim_{k, L \rightarrow \infty} E_{k}^L.
$$ \nomenclature{$PMI$}{Persistente Transinformation}
If the $PMI$ exists, it is enough to consider the iterated limits
$$
PMI = \lim_{L \rightarrow \infty} \lim_{k \rightarrow \infty} E_{k}^L =
\lim_{k \rightarrow \infty} \lim_{L \rightarrow \infty} E_{k}^L.
$$
\end{Def}

In the following we want to investigate this expression, which was proposed
first by Ball and collaborators in \cite{Bal10}. For stationary processes we can
write the persistent mutual information (if it exists) as
\begin{eqnarray}
PMI & = & \lim_{L \rightarrow \infty} \left ( H(L) - \lim_{k \rightarrow \infty} H(S_0^{L-1} | S_{-k - L+1}^{-k}) \right ) \nonumber \\
& = & \lim_{L \rightarrow \infty} \left ( 2 H(L) - \lim_{k \rightarrow \infty} H(S_0^{L-1} , S_{-k - L+1}^{-k}) \right ) \label{statPMI} \\
& = & \lim_{L \rightarrow \infty} \left ( H(L) - \lim_{k \rightarrow \infty} H(S_{k-1}^{k+L-2} | S_{-L}^{-1}) \right ) \nonumber \\
& = & \lim_{L \rightarrow \infty} \left ( 2 H(L) - \lim_{k \rightarrow \infty} H(S_{k-1}^{k+L-2} , S_{-L}^{-1}) \right ). \nonumber 
\end{eqnarray}
The last identities follow from the chain rule for the conditional entropy and
the stationarity of the process. Remark since the $PMI$ is assumed to exist it
is possible to exchange the limits.
\\

We now want to find a reasonable definition of persistent mutual information
for one-sided processes. A {\bf one-sided
process}\index{process!one-sided} is a stochastic process with indices
consisting only of positive or negative numbers, e.g. $\setN, \setZ_+$,
$\setR_+$ or $\setZ_-, \setR_-$. In order to achieve this we consider the
excess entropy of such a process. Because of the stationarity of the process we
can write the excess entropy as
$$
E = \lim_{L \rightarrow \infty} I(S_0^{L-1}; S_{-L}^{-1}) = \lim_{L
\rightarrow \infty} I(S_L^{2L-1}; S_0^{L-1}).
$$
Hence we obtain
$$
E_{t,\tau}^{L} = I(S_t^{t+L-1}; S_{-\tau -L+1}^{-\tau}) = I(S_{t+L}^{t+2L-1};
S_{-\tau +1}^{-\tau + L}),
$$
Because of the definition of one-sided processes we set $\tau = 1$ and obtain
the following definition.

\begin{Def}
Let a one-sided stochastic process with values in a finite alphabet $\kA$ be
given. The
\nt{persistent mutual information} of such a process is defined by
$$
PMI := \lim_{t, L \rightarrow \infty} I(S_{t+L}^{t+2L-1}; S_0^{L-1}).
$$
If $PMI$ exists it is enough to consider
$$
PMI = \lim_{L \rightarrow \infty} \lim_{t \rightarrow \infty} 
I(S_{t+L}^{t+2L-1}; S_0^{L-1}) = \lim_{t \rightarrow \infty} \lim_{L \rightarrow
\infty} I(S_{t+L}^{t+2L-1}; S_0^{L-1}).
$$
\end{Def}

Different to two-sided processes the reference point (which can be interpreted
as presence) moves to infinity. Like in the two-sided case we obtain
simpler expressions for stationary processes
$$
PMI = \lim_{L \rightarrow \infty} \left (H(L)-  \lim_{t \rightarrow \infty}
H(S_{t+L}^{t+2L-1} |S_0^{L-1} ) \right )= \lim_{L \rightarrow \infty} \left (2
H(L)-  \lim_{t \rightarrow \infty} H(S_{t+L}^{t+2L-1}, S_0^{L-1} ) \right ),
$$
where it is again allowed to change the limits if $PMI$ exists.

\begin{Bem}
Remark that both $PMI$-expressions are also defined for nonstationary
stochastic processes. In this paper we only consider stationary processes. In
any case the existence of $PMI$ is a priori not clear. Nevertheless we can show
that it exists for Markov processes of finite order or for periodic processes
(see Section \ref{sec.explrepr}).
\end{Bem}

\section{Necessary Conditions for Existence}

From the definition of $PMI$ it is not clear if the limits exist. In this
section we assume that the double sequence $(E_k^L)_{k,L \in \setN}$ converge
and hence the $PMI$ exists. We investigate some necessary conditions for the
existence of $PMI$, to be precise we investigate what can be deduced from the
existence of iterated limits
\begin{equation} \label{ExEquation}
\lim_{L \rightarrow \infty} \lim_{k \rightarrow \infty} E_k^L =  \lim_{L
\rightarrow \infty} \left ( 2 H(L) - \lim_{k \rightarrow \infty} H(S_0^{L-1} ,
S_{-k - L+1}^{-k}) \right )
\end{equation}
for a corresponding stochastic process. 

We consider two-sided stationary processes and
consider the inner limit of (\ref{ExEquation})
\begin{eqnarray*}
&& \lim_{k \rightarrow \infty} H(S_0^{L-1}, S_{-k-L+1}^{-k}) \\
& & = - \sum_{\sigma, \xi \in \kA^L} \lim_{k \rightarrow \infty} \left ( 
\Pr(S_0^{L-1} =\sigma, S_{-k-L+1}^{-k} = \xi) \log( \Pr(S_0^{L-1} = \sigma,
S_{-k-L+1}^{-k} = \xi) ) \right ).
\end{eqnarray*}

If this limit exist then the limit of the induced probability distribution also
exist
$$
\lim_{k \rightarrow \infty} \Pr(S_0^{L-1} = \sigma, S_{-k-L+1}^{-k} = \xi) =
\lim_{k \rightarrow \infty} P(\{\omega \in \Omega: S_0^{L-1} (\omega) = \sigma,
S_{-k-L+1}^{-k} (\omega) = \xi\}),
$$
with $\sigma, \xi \in \kA^L$.
This is a limit in the space of all probability distributions on $\kA^L$ which
we denote as $\kM(\kA^L)$. We introduce a topology on
$\kM(M)$\nomenclature{$\kM(M)$}{Space of all probability distributions on $M$}.
With $C(M)$\nomenclature{$C(M)$}{Space of all continuous functions on $M$} we
denote the space of all continuous functions $f : M \rightarrow \setR$.

\begin{Def} \label{weakstarTopology}
The \nt{weak* topology} on $\kM(M)$ is the smallest topology, such that for $\mu
\in \kM(M)$ and $f\in C(M)$ every map $\kM(M) \rightarrow \setC$ with $\mu
\mapsto \int_M f d\mu$ is continuous. A basis is given by 
$$
V_\mu (f_1, \dots, f_k; \epsilon) = \left \{\nu \in \kM(M) : \left | \int f_i d
\nu - \int f_i d \mu \right | < \epsilon , 1 \le i \le k \right \},
$$ 
with $\mu \in \kM(M), k \ge 1, f_i \in C(M)$ and $\epsilon > 0$.
\end{Def}

With this definition we can understand the limit above as a weak limit with
respect to this topology.
\begin{Def}[\cite{Bil68}]
A sequence $(P_n)_{n \in \setN}$ in $\kM(M)$ {\bf converges
weak*}\index{weak* convergence} to $P \in \kM(M)$, if for all $f \in C(M)$ it
holds that
$$\int_M f dP_n \stackrel{n \rightarrow \infty}{\rightarrow} \int_M f dP.$$
\end{Def}

The Portmanteau Theorem gives a series of equivalent characterizations of the
weak*-convergence.
\begin{Satz}[Portmanteau Theorem]
Let $P_n, P \in \kM(M)$ and $(M, \kF, P_n)$, \\
$(M, \kF, P)$ be probability spaces for $n \in \setN$. Then the following is
equivalent
\begin{itemize}
	\item[(i)] $P_n$ is weak* convergent to $P$.
	\item[(ii)] $\displaystyle \lim_{n \rightarrow \infty} \int_M f dP_n = \int_M
f dP$ for all $f \in C(M)$.
	\item[(iii)] $\displaystyle \limsup_{n \rightarrow \infty} P_n(F) \le P(F)$
for all closed sets $F \in \kF$.
	\item[(iv)] $\displaystyle \liminf_{n \rightarrow \infty} P_n(G) \ge P(G)$
for all open sets $G \in \kF$.
	\item[(v)] $\displaystyle \lim_{n \rightarrow \infty} P_n(A) = P(A)$ for all
sets $A \in \kF$ with $P(\partial A) = 0$, where the border of $A$ is denoted as
$\partial A$.
\end{itemize}
\end{Satz}
\begin{Proof}
See \cite{Bil68} Chapter 1.2.
\end{Proof}

In particular the last equivalence show that the weak* convergence can be
understood as pointwise convergence in our case, since for all sets $\sigma \in
\kA^L$ it holds $\partial \sigma = \emptyset$. A first answer to the
question when the distributions $\Pr(S_{-k-L+1}^{-k})$ have a limit with respect
to weak* convergence give the following Proposition, which is a version of
Proposition 5.5 in \cite{Bil68}.

\begin{Satz}
Let $P \in \kM(M)$, furthermore let $S_k : \Omega \rightarrow \kA$ be a
sequence of measureable mappings, which converge pointwise to a mapping $S:
\Omega \rightarrow \kA$ $P$-almost everywhere. Then it holds that
$$P(S_k^{-1}) \stackrel{k \rightarrow \infty}{\longrightarrow} P(S^{-1}),$$
w.r.t. weak*-topology.
\end{Satz}
\begin{Proof}
We show item $(iv)$ in the Portmanteau Theorem. To do that we define for an
open set $G \subset \kA$
$$
T_k := \bigcap_{i \ge k} S_i^{-1}(G) = \bigcap_{i \ge k} \left \{\omega:
S_i(\omega) \in G \right \}.
$$
Let $N$ be a set of measure zero w.r.t. $P$, containing those points for which
$S_k$ does not converge pointwise to $S$. It holds that
$$
\displaystyle S^{-1}(G) \subset \bigcup_{k} \mathring{T_k} \cup N.
$$
Because of $P(N)=0$ it follows that $\displaystyle P \left (S^{-1}(G) \right )
\le P \left (\bigcup_k \mathring T_k \right )$. Furthermore $\mathring T_k
\subset \mathring T_{k+1}$. For $\epsilon > 0$ and $k$ choosen large enough we
obtain
$$
P(S^{-1} (G)) \le P(\mathring T_k) + \epsilon.
$$
With $\mathring T_k \subset S_k^{-1}(G)$ it follows that $P(\mathring T_k) \le
P(S_k^{-1}(G))$. Putting things together we obtain
$$
P(S^{-1}(G)) \le P(\mathring T_k) + \epsilon \le P(S_k^{-1}(G)) + \epsilon.
$$
Since $\epsilon$ is arbitrary and the left-handside is not depending on $k$ we
get
$$
P(S^{-1}(G)) \le \liminf_{k \rightarrow \infty} P(S_k^{-1}(G)).
$$
\end{Proof}

\begin{Bem}
With the same argument as in the proof above one can show for the same
assumptions the convergence of a finite-length block of random variables
$$
P(S_k^{-1}, S_{k+1}^{-1}, \dots, S_{k+L-1}^{-1}) \stackrel{k \rightarrow
\infty}{\rightarrow} P(S^{-1}),
$$
w.r.t. weak*-topology. Furthermore one can extend the result to
joint distributions of different random variables
$$
P(S_0^{-1}, \dots, S_{L-1}^{-1}, S_k^{-1}, \dots, S_{k+L-1}^{-1}) \stackrel{k
\rightarrow \infty}{\rightarrow} P(S_0^{-1}, \dots, S_{L-1}^{-1}, S^{-1}),
$$
w.r.t. weak*-topology.
\end{Bem}

To fulfill the assumptions of the last proposition, the random variables of a
stochastic process need to converge almost everywhere. If the set of all points
$\omega \in \Omega$ for which
\begin{equation} \label{leftConvAssumption}
S_{-k} (\omega) \stackrel{k \rightarrow \infty}{\rightarrow} S(\omega),
\end{equation}
not hold, is a set of measure zero w.r.t. $P$, then the limit of the
distributions exist
$$
\lim_{k \rightarrow \infty} \Pr (S_0^{L-1} = \sigma , S_{-k-L+1}^{-k} = \xi) =
\Pr(S_0^{L-1} = \sigma, S = \xi_1).
$$
Hence the following limit exists
$$
\lim_{k \rightarrow \infty} H(S_0^{L-1}, S_{-k-L+1}^{-k}) = H(S_0^{L-1},S).
$$

With that we have shown the following proposition.

\begin{Satz} \label{statPMIExzess}
Assume that $PMI$ exists for a stationary stochastic process
$\overleftrightarrow{S}$ and the process
fulfill the convergence condition (\ref{leftConvAssumption}) $P$ a.e., then the
$PMI$ is the mutual information-version of the "excess entropy" of the
following stochastic process
\begin{equation} \label{NewProcess}
\tilde S = (\ldots, S, \ldots, S, S_{0}, \ldots, S_{L-1}, \ldots),
\end{equation}
to be precise $(\tilde S_t)_{t \in \setZ}$ with $\tilde S_t := S$ for all $t<0$
and $\tilde S_t := S_t$ for $t \ge 0$.
In general $(\tilde S_t)_{t \in \setZ}$ is not stationary and it holds that
$$
PMI = \lim_{L \rightarrow \infty} (2 H(L) - H(S_0^{L-1}, S)) = \lim_{L
\rightarrow \infty} I(S_0^{L-1}; S) = I(\overrightarrow{S};S).
$$
If the process $\overleftrightarrow{S}$ is a one-sided stationary process and
if we assume that
$S_t(\omega) \stackrel{t \rightarrow \infty}{\rightarrow} S(\omega)$ $P$-a.e., then the same result holds
\begin{equation}
PMI = \lim_{L \rightarrow \infty} (2 H(L) - H(S, S_0^{L-1})) = \lim_{L \rightarrow \infty} I(S; S_0^{L-1}) = I(S; \overrightarrow{S}). \tag*{$\Box$}
\end{equation}
\end{Satz} 

\begin{Bem}
If the process (\ref{NewProcess}) in Proposition \ref{statPMIExzess} is 
stationary it holds for two-sided and one-sided processes that $PMI=H(S)$.
\end{Bem}

The $PMI$ is the mutual-information based version of the excess entropy of a 
process with constant past (and with constant future in the one-sided case). In
the
original process this constant past is located very far in the past. The $PMI$
can thus be understood as the amount of information which is communicated from a
very far past to the future. In this sense the $PMI$ represents a kind of memory
which is permanently stored in the process for all times. Thus the $PMI$ can be
considered as an inherent measure of the system complexity.


\begin{Bem}
Note that because of
$$0 \le H(L) \le L \log(|\kA|)$$ 
and 
$$0 \le H(S_t^{t+L-1}, S_{-\tau-L+1}^{-\tau}) \le 2 L \log (|\kA|),$$
the convergence at $t, \tau \rightarrow \infty$ leads to a unique limit or to a
certain number of accumulation points. If the expression $H(S_t^{t+L-1},
S_{-\tau-L+1}^{-\tau})$ is monotonic increasing or decreasing w.r.t. $t$ or
$\tau$, a limit exists. Proposition \ref{statPMIExzess}
explains the $PMI$ only for the special class of processes for which
(\ref{ExEquation}) exists.
\end{Bem}

\section{Relation to Statistical Complexity}

We now pick up the sketched ideas in \cite{Bal10}, to express the $PMI$ with 
so called {\em causal states}. In particular one can show that the
{\em statistical complexity} (internal entropy of the causal states) is an upper
bound for the
$PMI$. In the rest of this section we assume that the $PMI$ exist. We start with
introducing time-indexed causal states. We consider shifted blocks of random
variables
$$
\overleftarrow{S}_\tau := \left (S_{-\tau-t} \right )_{t \in \setN}, \qquad
\overrightarrow{S}_\tau := \left (S_{\tau + t} \right )_{t \in \setN_0},
$$
for $\tau \in \setN_0$. The sets of realisations\footnote{For every $\omega \in
\Omega$ the mapping $R_\omega: t \mapsto S_{-t}(\omega)$ is called a {\rm
realisation} of the process $\overleftarrow{S}$. The set of all realisations is
defined as $\overleftarrow{\mathbf{S}}:=\{ (R_\omega(t))_{t \in \setN}: \omega
\in \Omega\}$.} are denoted by $\overleftarrow{\mathbf{S}}_\tau \subset
\overleftarrow{\mathbf{S}},
\overrightarrow{\mathbf{S}}_\tau \subset \overrightarrow{\mathbf{S}}$ and the
sub-$\sigma$-algebras which are generated by cylinder sets are denoted with
$\kC_{\tau, -\setN}
\subset \kC_{-\setN}, \kC_{\tau, \setN_0} \subset \kC_{\setN_0}$. On the set 
$\kA^\setN$ of all shifted past trajectories of the process
$\overleftrightarrow{S}$ we define an equivalence relation
$$
\overleftarrow{s} \sim \overleftarrow{s}' : \Leftrightarrow
\Pr(\overrightarrow{S} = \overrightarrow{s} \mid \overleftarrow{S}_\tau =
\overleftarrow{s}) = \Pr(\overrightarrow{S} = \overrightarrow{s} \mid
\overleftarrow{S}_\tau = \overleftarrow{s}'), \quad \forall \overrightarrow{s}
\in \kC_{\setN_0},
$$
where $\overleftarrow{s},\overleftarrow{s}' \in \kA^\setN$ and
$\Pr(\overrightarrow{S} = \overrightarrow{s} \mid \overleftarrow{S}_\tau =
\overleftarrow{s})$ 
is a regular version of the conditional expectation. The equivalence classes
$$
S^+_\tau(\overleftarrow{s}) := \{\overleftarrow{s}' \in \kA^\setN :
\overleftarrow{s}' \sim_\tau \overleftarrow{s}\} \subset \kA^\setN
$$
of this relation are called {\em shifted causal states}. The set of all shifted
causal states is denoted by $\kS_\tau^+ := \{S_\tau^+(s) \mid s \in
\kA^\setN\}$.

In the same sense we define (future) shifted causal states
$S_\tau^-(\overrightarrow{s})$ and $\kS_\tau^-$ (we only have to change the rule
of past and future trajectories $\overleftarrow{s}$ and $\overrightarrow{s}$).

We are only considering stationary stochastic processes with a finite set of 
shifted causal states $\kS_\tau^+ = \{S^+_1, \ldots, S^+_n\}$ and $\kS_\tau^- =
\{S^-_1, \ldots, S^-_m\}$. Given a past observation of infinite length
$s_{-\infty}^t \in \kA^{\setZ}$ at time $t \in \setZ$ using stationarity we
identify this shifted past with a shifted causal state
$S_\tau^+(\sigma^{-t-\tau-1}(s_{-\infty}^t)) \in \kS_\tau^+$. Together with the
next symbol $s_{t+1}$ generated by the process the next shifted causal state
$S_\tau^+(\sigma^{-t-\tau-2}(s_{-\infty}^t s_{t+1})) \in \kS_\tau^+$ is uniquely
determined and the shifted causal states are Markov \cite{Sha01, Loe10}. We
define the Markov kernels between two shifted causal states $S_i^+, S_j^+ \in
\kS_\tau^+$ emitting an output symbol $r \in \kA$ for any $t\in \setZ$ as
follows
\begin{eqnarray*}
T_{\tau,i,j}^{+(r)} & := & T(S_i^+)(S_j^+, r) \\
& = & \Pr \left (S(\sigma^{-t-\tau-2}(s_{-\infty}^t s_{t+1})) = S_j^+ \, {\rm
and} \, S^+_{t+1} = r \, \left | \, S(\sigma^{-t-\tau-1}(s_{-\infty}^t)) = S_i^+
\right. \right ).
\end{eqnarray*}
The set of transition matrices is denoted with $T^+_\tau :=
\left \{ \left (T_{\tau,i,j}^{+(r)} \right )_{i,j=1}^n : r \in \kA \right \}$.
The probability of a shifted causal state $S_i^+ \in \kS_\tau^+$ is denoted by $p_i^+ := P(S_i^+)$.
The ordered pair $M^+_\tau:=(T^+_\tau, (p_1^+, \ldots, p_n^+))$ is called {\em
shifted (past) $\epsilon$-machine}. In the same way we can define a {\em shifted
(future) $\epsilon$-machine} $M^-_\tau:=(T^-_\tau, (p_1^-, \ldots, p_m^-))$.

The shifted $\epsilon$-machines has internal state entropies
$$C^+_{P,\tau} := H(\kS_\tau^+) = - \sum_{j=1}^n p_j^+ \log p_j^+,$$
and
$$C^-_{P,\tau} := H(\kS_\tau^-) = - \sum_{j=1}^m p_j^- \log p_j^-,$$
which are also known as (shifted) {\em statistical complexities} \cite{Gra86, Sha01}. 

We can write the $PMI$ as follows.
\begin{Satz} \label{PMIViaProcess}
Assume that $PMI$ exists for a stationary stochastic process, then it
holds that
$$
PMI = \lim_{\tau \rightarrow \infty}
I(\overleftarrow{S};\overrightarrow{S}_\tau).
$$
\end{Satz}
\begin{Proof}
Since $PMI$ exists we can change the limits and write them as
\begin{equation} \label{PMIEqu}
PMI = \lim_{\tau \rightarrow \infty} \lim_{L \rightarrow \infty} I(S_{-L}^{-1}; S_{\tau-1}^{\tau+L-2}).
\end{equation}
We can decompose the limits in (\ref{PMIEqu}) into two independent limits and
get with Proposition \ref{InfoProperties} (iii) applied two times
$$
PMI = \lim_{\tau \rightarrow \infty} \lim_{L \rightarrow \infty}
I(S_{-L}^{-1}; S_{\tau-1}^{\tau+L-2}) = \lim_{\tau \rightarrow \infty}
I(\overleftarrow{S}; \overrightarrow{S}_\tau).
$$
\end{Proof}

Similar to the fact that the excess entropy can be expressed via causal 
states (see Proposition \ref{ExcessEntKausal} and \cite{Ell09}), we can also
express the $PMI$ via shifted causal states.

\begin{Satz} \label{PMIKausalStates}
Assume the $PMI$ of a stationary stochastic process exists. Then we can write 
the $PMI$ as
$$
PMI = \lim_{\tau \rightarrow \infty} I (\kS_0^+ ; \kS^-_{\tau}) = \lim_{\tau
\rightarrow \infty} I(\kS_\tau^+; \kS^-_0).
$$
\end{Satz}
\begin{Proof}
We take $\overrightarrow{S}_\tau$ instead of $\overrightarrow{S}$ and 
$\kS^+_0, \kS^-_\tau$ instead $\kS^+, \kS^-$, then the proof is with Proposition
\ref{PMIViaProcess} analogous to the proof of Proposition
\ref{ExcessEntKausal}. We obtain the second equality with the symmetry of the
mutual information and the stationarity of the stochastic process.
\end{Proof}

With this expression we get the following inequalities.
\begin{Korollar}
The statistical complexity is an upper bound for the $PMI$, if it exists,
$$
PMI \le C_P^-, \qquad PMI \le C_P^+,
$$
with equality if and only if $\displaystyle \lim_{\tau \rightarrow \infty}
H(\kS_0^+ |\kS_\tau^-) = 0$ or $\displaystyle \lim_{\tau \rightarrow \infty}
H(\kS^{-} | \kS_\tau^+) = 0$.
\end{Korollar}
\begin{Proof}
With Proposition \ref{PMIKausalStates}, the stationarity and the definition of the statistical complexity we get
$$
PMI = H(\kS_0^+) - \lim_{\tau \rightarrow \infty} H(\kS_0^+ | \kS_\tau^-) \le
C_P^+.
$$
With the symmetry of the mutual information we get $PMI \le C_P^-$.
\end{Proof}

\begin{Korollar} \label{StatPMIRel}
It holds that
$$
C_P^+ = PMI + \lim_{\tau \rightarrow \infty} H(\kS^+ | \kS_\tau^-), \qquad C_P^-
= PMI + \lim_{\tau \rightarrow \infty} H(\kS^- | \kS_\tau^+),
$$
furthermore we have the following inequalities
$$
C_P^+ \ge \lim_{\tau \rightarrow \infty} H(\kS^+ | \kS_\tau^-), \qquad C_P^-
\ge\lim_{\tau \rightarrow \infty} H(\kS^- | \kS_\tau^+).
$$
\end{Korollar}
\begin{Proof}
With Proposition \ref{PMIKausalStates}, the definition of statistical 
complexity and the symmetry of mutual information the equalities follow.
Because of $PMI \ge 0$ we obtain the inequalities.
\end{Proof}

\begin{Bem}
If $PMI = 0$, then we get with Corollary \ref{StatPMIRel} that 
$$
C_P^+ = H(\kS^+) = \lim_{\tau \rightarrow \infty} H(\kS^+ |\kS_\tau^-).
$$
This is the case if and only if 
$\displaystyle \lim_{\tau \rightarrow \infty} \kS_\tau^-$ and $\kS^+$ are 
stochastic independent. This means that the causal states in the very far past
are independent from the causal states in the future. 
\end{Bem}

\section{Relation to Excess Entropy}

In this section we want to find relations between $PMI$ and excess entropy. One 
might expect that the persistent mutual information coincide with the
excess entropy as soon as the structure of the past coincide with the structure
of the future. The next proposition shows that this is indeed the case for
processes with zero metric entropy. The metric entropy is defined as the
following limit $h_P:= \lim_{L \rightarrow \infty} \frac{H(L)}{L}$ and exist
for all stationary processes.

\begin{Satz} \label{PMIEqualE}
Assume that the excess entropy $E$ and $PMI$ exists for a stationary stochastic 
process, then it holds that
$$
PMI = E \iff h_P = 0 \iff \lim_{\tau \rightarrow \infty} H(\kS^+ | \kS_\tau^-) =
0 \; {\rm and} \; H(\kS^+ | \kS^-) = 0.
$$
\end{Satz}
\begin{Proof}
We prove the first equivalence. \\
$\Rightarrow$: Since $E$ is finite we get with Proposition \ref{SatzConv} that 
$H(L) \sim E + L h_P$ as $L \rightarrow \infty$. Furthermore we get
\begin{eqnarray}
PMI & = & \lim_{L \rightarrow \infty} (2 H(L) - \lim_{k \rightarrow \infty} H(S_0^{L-1}, S_{-k-L+1}^{-k})) \label{FirstEqu} \\
& = & \lim_{L \rightarrow \infty} (2 E + 2 L h_P - \lim_{k \rightarrow \infty} H(S_0^{L-1}, S_{-k-L+1}^{-k})).
\end{eqnarray}
Hence we get with $PMI = E$
$$
\lim_{L \rightarrow \infty} \left ( \lim_{k \rightarrow \infty} H(S_0^{L-1}, S_{-k-L+1}^{-k}) - 2 L h_P \right ) = 2E-PMI = E =  \lim_{L \rightarrow \infty} (H(L) - H(S_0^{L-1} | S_{-L}^{-1})).
$$
Since $\displaystyle 1 \le \frac{\lim_{k \rightarrow \infty} H(S_0^{L-1} ,
S_{-k-L+1}^{-k} )}{H(L)}$ and $\displaystyle \frac{H(L)}{\lim_{k \rightarrow
\infty} H(S_0^{L-1} , S_{-k-L+1}^{-k} )} \le 1$, it follows that
$$
\lim_{k \rightarrow \infty} H(S_0^{L-1} , S_{-k-L+1}^{-k} ) \sim H(L) \; \; {\rm
as} \; L \rightarrow \infty,
$$
which leads with (\ref{FirstEqu}) to
$$
PMI = \lim_{L \rightarrow \infty} H(L) = E.
$$
Finally this implies because of Proposition \ref{SatzConv} that $h_P = 0$. \\
$\Leftarrow$: Due to $h_P = 0$ it holds that $\displaystyle E = \lim_{L \rightarrow \infty} H(L)$. Furthermore it follows that
$$
H(2L + k) \ge H(S_0^{L-1}, S_{-k-L+1}^{-k}) \ge
H(L),
$$
using $H(2L + k) \stackrel{k \rightarrow \infty}{\rightarrow} E$ and
$H(L) \stackrel{L \rightarrow \infty}{\rightarrow} E$
leads to
$$
\lim_{L \rightarrow \infty} \lim_{k \rightarrow \infty} H(S_0^{L-1}, S_{-k-L+1}^{-k}) = E.
$$
Together we get
$$
PMI = \lim_{L \rightarrow \infty} (2 H(L) - \lim_{k \rightarrow \infty} H(S_0^{L-1}, S_{-k-L+1}^{-k}) ) = 2E - E = E.
$$
The second equivalence follows with Corollary \ref{statCompExzessEnt}, Corollary \ref{StatPMIRel} and simple transformations.
\end{Proof}

More generally we can show that the $PMI$ is bounded from above by the 
excess entropy.
\begin{Satz} \label{PMIlessE}
Assume that the $PMI$ exists for a stationary stochastic process then it holds 
$$PMI \le E.$$
\end{Satz}
\begin{Proof}
With Propositions \ref{PMIKausalStates}, \ref{ExcessEntKausal} and the rule
$H(X|Y) \le H(X)$ for two random variables $X,Y$, we get

\begin{eqnarray*}
PMI = \lim_{\tau \rightarrow \infty} I(\kS^-; \kS^+_\tau) & = & \lim_{\tau
\rightarrow \infty} (H(\kS^-) - H(\kS^- | \kS_\tau^+)) \\
& = &\lim_{\tau \rightarrow \infty} (H(\kS^-) - H(\kS^-
|\overleftarrow{S}_{\tau})) \\
& = &\lim_{\tau \rightarrow \infty} \lim_{L \rightarrow \infty} (H(\kS^-) -
H(\kS^- |S_{-\tau-L+1}^{-\tau-1})) \\
& \le & \lim_{\tau \rightarrow \infty} \lim_{L \rightarrow \infty} (H(\kS^-) -
H(\kS^- | S_{-\tau-L+1}^{-1})) \\
& = & H(\kS^-) - H(\kS^- | \overleftarrow{S}) \\
& = & H(\kS^-) - H(\kS^- | \kS^+) \\
& = & I(\kS^-; \kS^+) = E.
\end{eqnarray*}
\end{Proof}


\begin{Bem}
The $PMI$ do not care about some random variables which are considered by the
excess entropy. Proposition \ref{PMIlessE} tells us that $PMI$ forgets this
information and the excess entropy use the full
information available from the realisations of the process. In this sense the
$PMI$ is a coarser complexity measure than the excess entropy. With that we get
a graduation of the considered complexity measures from a coarse to a fine one,
i.e. $$PMI \le E \le C_P^+.$$
\end{Bem}



\section{Explicit Representations} \label{sec.explrepr}
We show a series of explicit representations of the $PMI$ for simple processes.
First we consider a simple case in which the metric entropy vanishes and
periodicity is part of the process, i.e. periodic processes\footnote{A process
is called {\em periodic} with period $L$ if $S_t = S_{t+L}$ for all $ t\in
\setZ$ and $S_t \neq S_{t+k}$ for $k < L$.}. For that case the
following corollary of Proposition \ref{PMIEqualE} give us the result.

\begin{Korollar} \label{PMIPeriodSatz}
Let a stationary periodic process with period $L$ be given. Then the persistent 
mutual information amounts to 
$$PMI = H(L),$$
in particular it holds $PMI = E$.
\end{Korollar}
\begin{Proof}
With Proposition \ref{PMIEqualE} and the fact that $h_P = 0$ hold for periodic 
processes, the claim follows with the fact that $E=H(L)$ for periodic processes.
We show an additional more elementary proof of the
corollary which shows the result for an iterated limit like (\ref{ExEquation})
and which show the existence of the $PMI$ for $L$-periodic processes. Because
the process is $L$-periodic it holds that
$$
\Pr(S_1, \ldots, S_{L+1}) = \Pr (S_1, \ldots, S_L),
$$
and $H(M) = H(L)$ for $M > L$.
Consider $M=L+1$ and the joint probability distribution, then one obtains 
with $(-\tau-t-M+1)\mod L = k$
\begin{eqnarray*}
&\Pr& (S_1^{M} = \sigma_1^M, S_{-\tau - t -M+1}^{-\tau -t} = \xi_1^M) \\
& = & P(\{ \omega: S_1^{M}(\omega) = \sigma_1^M \wedge S_{-\tau -t -M+1}^{-\tau -t}(\omega) = \xi_1^M\}) \\
& = & \left \{ \begin{array}{l l} P(\{\omega: S_1^{L}(\omega) = \sigma_1^L
\wedge S_{k}(\omega) = \xi_{1} \wedge \ldots \wedge S_{k} (\omega) =
\xi_{M}\}), & {\rm if} \; \sigma_1 = \sigma_{L+1}, \\ 0, & {\rm else,}
\end{array} \right. \\
& = & \left \{ \begin{array}{l l} \Pr(S_1^L = \sigma_1^L), & {\rm if} \;
\sigma_1 = \sigma_{L+1}, \xi_{1} = \xi_M = \sigma_k, \ldots, \xi_{M-1} =
\sigma_{k-1}, \\ 0, & {\rm else.} \end{array} \right.
\end{eqnarray*}
With the definition of $H(S_0^{M-1}, S_{-\tau -t -M+1}^{-\tau -t})$ it holds 
for $M > L$ that
$$
H(S_0^{M-1} , S_{-\tau -t-M+1}^{-\tau -t}) = H(L).
$$
Finally we get for the persistent mutual information
$$
PMI = \lim_{M \rightarrow \infty} \left (2 H(M) - \lim_{t \rightarrow \infty} \lim_{\tau \rightarrow \infty} H(S_0^{M-1}, S_{-\tau -t-M+1}^{-\tau-t}) \right ) = H(L).
$$
\end{Proof}

For Markov-processes the $PMI$ vanishes, since the dependencies between the 
past and future blocks disappear in finite time.

\begin{Satz} \label{MarkovPMI}
Let a Markov-process of order $R$ be given\footnote{A stochastic process is
called Markov of order $R$ if for all $t\ge t_n > \ldots > t_0 \ge 0$ it holds
that $\Pr(S_t|S_{t_n}, \ldots, S_{t_0})=\Pr(S_t|S_{t_n}, \ldots, S_{t_n -R+1})$
for all $n \ge R$.}. Then it holds that
$$
PMI = 0.
$$
\end{Satz}
\begin{Proof}
With the Markov-property and the abbreviation $\tilde S_k := S_{-k-L+1}^{-k}$
it follows for $-k < R$
\begin{eqnarray*}
H(S_0^{L-1} \mid \tilde S_k) & = & - \sum_{\sigma, \xi \in \kA^L} \Pr(S_0^{L-1}
= \sigma , \tilde S_k = \xi) \log (\Pr(S_0^{L-1} = \sigma \mid \tilde S_k = \xi)
) \\
& = & - \sum_{\sigma, \xi \in \kA^L} \Pr(S_0^{L-1} = \sigma \mid \tilde S_k =
\xi) \Pr(\tilde S_k = \xi) \log (\Pr(S_0^{L-1} = \sigma \mid \tilde S_k = \xi) )
\\
& \stackrel{-k < R}{=} & -\sum_{\sigma, \xi \in \kA^L} \Pr(S_0^{L-1} = \sigma)
\Pr(\tilde S_k = \xi) \log (\Pr(S_0^{L-1} = \sigma) ) \\
& = & H(L).
\end{eqnarray*}
Hence the persistent mutual information is
$$
PMI =\lim_{L \rightarrow \infty} \left (H(L) - \lim_{k \rightarrow \infty}
H(S_0^{L-1} \mid	 S_{-k-L+1}^{-k}) \right ) = \lim_{L \rightarrow \infty} \left
(H(L) - H(L) \right ) = 0.
$$
\end{Proof}

\section{Example Processes}

In the following we investigate concrete examples of stochastic processes and
calculate the $PMI$ for them.

\subsection{Independent, identical, distributed Process}
A stochastic process is called independent, identical distributed if the finite
dimensional distributions are independent and all distributions are equal, i.e.
if for finite times $t_1 < \ldots < t_n \in \setZ$ it holds that $\Pr(S_{t_1},
\ldots, S_{t_n}) = \Pr(S_{t_1}) \cdot \ldots \cdot \Pr(S_{t_n})$ and
$\Pr(S_{t_i}) =\Pr(S_{t_j})$ for all $i,j \in \{1,\ldots,n\}$.
The probability distributions are not depending on the time distance because they are identical distributed. Hence it holds that
$$
H(S_t^{t+L-1} , S_{-\tau -L+1}^{-\tau}) = H(S_0^{L-1}, S_{-L}^{-1}).
$$
Hence the $PMI$ coincide with the excess entropy $E$ by definition. Furthermore
we have
$$
\Pr(S_0^{L-1}, S_{-L}^{-1}) = \Pr(S_0^{L-1}) \Pr(S_{-L}^{-1}).
$$
With the definition of the mutual information we get for every $L \in \setN$
\begin{equation*}
I(S_0^{L-1}; S_{-L}^{-1} ) = \sum_{s_0^{L-1}, s_{-L}^{-1} \in \kA^L}
\Pr(S_0^{L-1} = s_0^{L-1}, S_{-L}^{-1} = s_{-L}^{-1}) \log \frac{\Pr(S_0^{L-1} =
s_0^{L-1}, S_{-L}^{-1} = s_{-L}^{-1})}{\Pr(S_0^{L-1} = s_0^{L-1})
\Pr(S_{-L}^{-1} =s_{-L}^{-1})} = 0,
\end{equation*}
and finally
$$
PMI = E = \lim_{L \rightarrow \infty} I(S_0^{L-1}; S_{-L}^{-1}) = 0. 
$$

\subsection{Thue-Morse Process}

The Thue-Morse sequence has been discovered the first time in 1851 by Prouhet as
a solution of the {\em Prouhet-Tarry-Escott-problem}. It was rediscovered in
1912 by Thue and 1921 by Morse in different settings. The sequence appears in many different mathematical fields and is well studied. This diversity leads
to a series of equivalent definitions of the sequence (there exists at least
ten different ways to define it). We choose a definition which is based on
substitutions and which is particulary easy (see Appendix
\ref{AppendixSpectral} for a short introduction to substitution systems). The
Thue-Morse sequence consists of a two symbol alphabet $\kA = \{0,1\}$ and is
constructed with the following substitution
$$
\kG: \kA \rightarrow \kA^2, \quad {\rm with} \; \; \kG (0) = 01, \; \; \kG(1) =
10.
$$
The Thue-Morse sequence is defined as the fixed point of $\kG$ with
$$
u := \kG^\infty(0) = \lim_{t \rightarrow \infty} \kG^t (0) = 011010011001\ldots =  \kG(u).
$$
The stochastic process which generates the Thue-Morse sequence is called {\em
Thue-Morse process}. We want to calculate the $PMI$ for that process. The used
probability measure is the counting measure. To be precise for a block of
symbols $F$ of length $n$ in $u$ we define the counting measure as follows
$$
\Pr(F) := \lim_{j \rightarrow \infty} \frac{1}{j} \left | \{n < j : u_n \ldots
u_{n+|F|-1} = F\} \right |.
$$
We denote a block of symbols of length $n$ in $u$ as {\em factor}. We can
calculate the frequency and hence the probability of a factor in $u$ with
spectral analytical methods \cite{Que87}. One can show that this substitution
system is uniquely ergodic\footnote{A substitution system is called uniquely
ergodic, if there exist a unique invariant probability measure.} (see Appendix
\ref{AppendixSpectral} for a more detailed discussion). Hence the Thue-Morse
process is a stationary stochastic process and we can calculate the $PMI$. For
that we follow \cite{Ber94} to calculate the frequencies of factors in $u$. The
key for that calculation is the following lemma which is proved in \cite{Fog08}
and in \cite{Que87}.

\begin{Lemma} \label{RecognizabilityLemma}
Every factor $F$ of length $n>4$ in $u$ has an unique preimage (up to
possibly appearing border terms) w.r.t. the substitution $\kG$.
\end{Lemma}
\begin{Proof}
First we show that the Thue-Morse sequence does not contain blocks of symbols
with more than two identical consecutive symbols. Otherwise if the block $000$
exists in $u$, there need to exist a symbol $a \in \kA$ with $\kG(a) = 00$. But
with the definition given above that is not the case and thus all blocks with
zeros of greater length than two are also excluded. Analogous one see that also
no blocks of ones with length greater than two appear in $u$. With the same
argument one can show that there are no blocks of the form $01010$ or $10101$
in $u$, since the preimage of such blocks would be $000$ or $111$. That means
that in $F$ at least one of the blocks $00$ or $11$ appear. Split $F$ into
blocks of length two such that none of these smaller blocks is $00$ or $11$
(possibly there remain some boundary blocks of length one). This splitting is
with the remarks above uniquely determined and gives us the unique preimage of
$F$.
\end{Proof}

This fundamental property is also known as {\em recognizability-property} of a
substitution system, see \cite{Que87} for more details. In \cite{Dek92} Dekking
shows the following proposition.

\begin{Satz} \label{TMProb}
Factors of length $n \ge 2$ in the Thue-Morse sequence with $2^{k} + 1 \le n \le
2^{k+1}$, $k \in \setN_0$ have the following frequencies
$$
\frac{1}{3 \cdot 2^k}, \quad \frac{1}{6 \cdot 2^k}.
$$
Factors of length $1$ appear with frequency $\frac{1}{2}$.
\end{Satz}
\begin{Proof}
We prove the claim by induction on $k$.\\
For $k=0$ and $k=1$ we have that $n=2$ or $n=3, 4$ respectively and calculate
the frequencies like in \cite{Que87} via a spectral analysis (see Appendix
\ref{AppendixSpectral} for details) and get $\frac{1}{3}$,$\frac{1}{6}$ or
$\frac{1}{3 \cdot 2}, \frac{1}{6 \cdot 2}$ respectively. Assume now that the
claim is proved for a $k$ and we want to show the induction step from $k$ to
$k+1$. It holds that $2^{k+1}+1 \le n \le 2^{k+2}$ and with Lemma
\ref{RecognizabilityLemma} a factor $F$ of length $n> 4$ have an unique preimage
$F'$. Since for a $p > n$ the frequency of $F$ in the first $2p$ letters of the
Thue-Morse sequence equals the frequency of $F'$ in the first $p$ letters (this
follows from the construction of the Thue-Morse sequence) for the frequency of
$F$ in $u$ it follows that
$$
\Pr(F) = \frac{\Pr(F')}{2}.
$$
With the induction assumption the claim follows.
\end{Proof}

If a factor can be continued by adding a letter to the right in at least two
different ways such that the continued symbol block is also a factor in $u$, we
call such a factor a {\em right specialfactor}. In our case this means that for
a right specialfactor $B$ the words $B0$ and $B1$ are also factors in $u$.
Dekking also proved the following Lemma \cite{Dek92}.
\begin{Lemma}	\label{rightSpecialLemma}
Let $F$ be a right specialfactor of length $n > 2$ in $u$ and $2^k+1 \le n \le
2^{k+1}$. Then $F$ has the frequency $\frac{1}{3 \cdot 2^k}$ and the right
extensions of $F$ have the frequency $\frac{1}{6 \cdot 2^k}$.
\end{Lemma}
From that the following important proposition can be derived, which gives us an
explicit expression for the first derivative of the block entropy.
\begin{Satz}[\cite{Ber94}]
For all $k \ge 1$ we have the following explicit expressions for the first
derivative of the block entropy $\triangle H(n) := H(n)-H(n-1)$
$$
\triangle H(n) = \frac{4}{3 \cdot 2^k}, \qquad {\rm if} \; \; 2^k+1 \le n \le 3
\cdot 2^{k-1},
$$
$$
\triangle H(n) = \frac{2}{3 \cdot 2^k}, \qquad {\rm if} \; \; 3 \cdot 2^{k-1} +1
\le n \le 2^{k+1}.
$$
\end{Satz}
\begin{Proof}
We use the abbreviation $\eta (x) := -x \log (x)$. The first derivative of the
entropy can be written as
$$
\triangle H(n) = \sum_{B \in S_n} \eta (\Pr(B0)) + \eta(\Pr(B1)) -\eta(\Pr(B)),
$$
where $S_n$ is the set of all right specialfactors of length $n$. The
cardinality of $S_n$ is given by the complexity function $p : \setN \rightarrow
\setN$. $p(n)$ is defined as the number of factors of length $n$ in $u$ and
so we have $|S_n |= p(n+1) - p(n)$. In \cite{deL89} the following property
of $p(n)$ for $u$ is shown
$$
p(n+1) - p(n) = 4, \qquad {\rm if} \; \; 2^k+1 \le n \le 3 \cdot 2^{k-1},
$$
$$
p(n+1) - p(n) = 2, \qquad {\rm if} \; \; 3 \cdot 2^{k-1} +1 \le n \le
2^{k+1}.
$$
Using Lemma \ref{rightSpecialLemma} we obtain for $2^k+1 \le n \le 3 \cdot
2^{k-1}$
$$
\triangle H(n) = (p(n+1) - p(n)) \left (2 \eta \left (\frac{1}{6 \cdot 2^k}
\right ) - \eta \left ( \frac{1}{3 \cdot 2^k} \right ) \right ) = \frac{4}{3
\cdot 2^k}.
$$
The claim for the case $3 \cdot 2^{k-1} +1 \le n \le 2^{k+1}$ follows in
an analogous way.
\end{Proof}

Thus the metric entropy vanishes $\displaystyle h_P = \lim_{n \rightarrow
\infty} \triangle H(n) = 0$. 
Since the Thue-Morse process is a one-sided process it is enough to consider
$H(L)$ and $H(S^{t+2L-1}_{t+L} , S_0^{L-1})$. With Proposition \ref{TMProb} we
get
$$
H(L) = - \sum_{\sigma \in \kA^L} \Pr (S_0^{L-1} = \sigma) \log \Pr (S_0^{L-1} = \sigma) \ge C \frac{p(L)}{L-1} \log (L-1),
$$
with a finite constant $C$. Furthermore one observe that due to Lemma 
\ref{RecognizabilityLemma} (which is essentially the recognizeability-property
of the substitution systems) a gap between two symbol blocks is uniquely
determined with the blocks of the border. More precisely that means that with
given borderblocks $S_{t+L}^{t+2L -1} = \sigma$ and $S_0^{L-1} = \xi$ there is
exactly one $\eta$, such that $S_0^{t+2L-1} = \xi \eta \sigma$ holds.
The same holds for the probabilities such that with $2^\tau+1 \le
2L+t-1 \le 2^{\tau+1}$ and Proposition \ref{TMProb} we get
\begin{eqnarray*}
H(S_{t+L}^{t+2L -1}, S_0^{L-1}) & = & - \sum_{\sigma, \xi \in \kA^L} \Pr(S_{t+L}^{t+2L-1} = \sigma, S_0^{L-1} = \xi) \log \Pr (S_{t+L}^{t+2L-1} =\sigma, S_0^{L-1} = \xi) \\
& = &- \sum_{\sigma, \xi \in S_L} \Pr(S_0^{t+2L-1} = \xi \eta \sigma) \log \Pr(S_0^{t+2L-1} = \xi \eta \sigma) \\
& \le & |\kA|^{2L} \frac{1}{3 \cdot 2^\tau} \log (6 \cdot 2^{\tau}).
\end{eqnarray*} 
Finally it follows that
$$
- H(S_{t+L}^{t+2L -1}, S_0^{L-1}) \ge -|\kA|^{2L} \frac{1}{3 \cdot 2^\tau} \log (6 \cdot 2^\tau) \stackrel{\tau \rightarrow \infty}{\rightarrow} 0,
$$
and thus
$$
I(S_{t+L}^{t+2L-1}; S_0^{L-1}) = (2 H(L) - H(S_{t+L}^{t+2L-1}, S_0^{L-1})) \ge
2H(L) - |\kA|^{2L} \frac{\log(6 \cdot 2^\tau)}{3 \cdot 2^\tau} \stackrel{\tau
\rightarrow \infty}{\rightarrow} 2H(L).
$$
Due to $2H(L) \stackrel{L \rightarrow \infty}{\rightarrow} \infty$ it holds that
$PMI = \infty$ (remark that one divergent sequence in the double sequence is
enough to derive the divergence of the double sequence). In particular one can
also show that $E = \infty$. There seems to exist a whole set of
further substitution processes for which the $PMI$ is infinite.

\subsection{Persistent Mutual Information for an one-dimensional
Ising-spinchain}

We calculate the $PMI$ for an one-dimensional Ising-spinchain. Due to the fact
that the spinchain is a Markov-process of first order it immediately follows
from Proposition \ref{MarkovPMI} that $PMI = 0$.

\begin{Bem}
Crutchfield et al. calculated in \cite{Cru97} and in \cite{Fel98} an explicit
expression of the excess entropy $E$ for that example, see also \cite{Gme10} for a detailed treatment. It turns out that
depending on the temperature the excess entropy attains a maximum at some
critical temperature and get close to zero for very low and very high
temperatures, see Figure \ref{fig.excessIsing}. 

\begin{figure}[H]
\begin{center}
\epsfig{figure=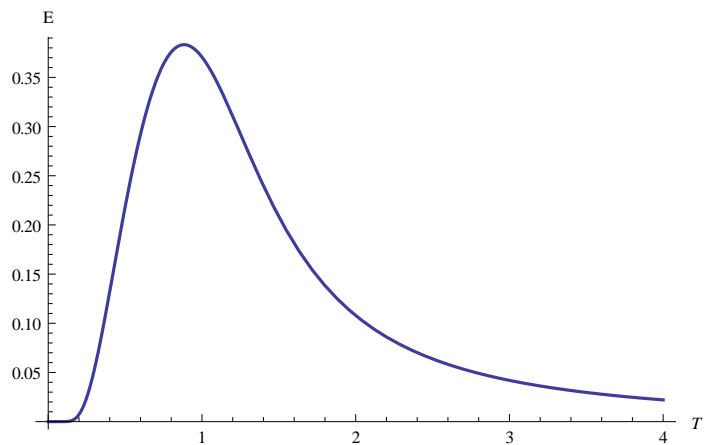}
\caption{Excess entropy for one-dimensional Ising model depending on the
temperature $T$.}
\phantomsection \label{fig.excessIsing}
\end{center}
\end{figure}

It is well known that in the
one-dimensional Ising model no phase-transition appears (we consider a
phase-transition as an example for {\it weak emergence}). Nevertheless the fact
that
$E$ attains a nontrivial expression in that case and $PMI$ is zero shows us that
$E$ seems to measure complexity structure at a too fine level and on the
first sight cannot distinguish between emergent structures and not emergent
structures. On the other hand the fact that $PMI$ is zero supports the intuition
that $PMI$ only detects emergent structures. To confirm this intuition we need
more concrete calculation examples, see Section \ref{chap:Emergenz} for a
detailed discussion.
\end{Bem}

\section{Emergence} \label{chap:Emergenz}
\thispagestyle{plain}

After defining mathematical measures for complexity we want investigate
their relation to emergent structures appearing in nature. In particular we try
to answer the question to what extend the introduced complexity measures and in
especially the $PMI$ are able to detect emergent structures. Before we start
with that we need to write down precisely what the term emergent structure and emergence means. In many works considering this topic this is
often a crucial part since the term emergence is often misunderstood and used
without a precise definition. There is a vast and confusing usage of the term
emergence in the literature for many situations which seem to have something in
common but differ at some point. Furthermore many people argue on an intuitive
level and do not define emergence in a precise way. A similar difficulty seems
to exist for the term complexity. There are a lot of papers concerning
complexity but often a clear mathematical definition is missing. However we try
in this section to give a clear description of emergence (at least we want
to define the meaning of the term in our sense). For that we start with a short
overview and go back to the roots of emergence which can be found in philosophy.

\subsection{Emergence - an Artificial Expression in Philosophy}
\label{PhilEmergentDef}

The term emergence is basically an artificial expression in philosophy which
is nowadays spreaded in many different scientific disciplines. The starting
point of emergent thinking goes back to Henry Lewes (1817-1878) and Stuart Mill
(1806-1873). The golden age of emergentism took place in the 1920s, mainly in
Great Britain. During this time many authors developed, independent of each
other, different theories of emergence. In particular the work ``The mind and
its place in nature'' \cite{Bro25} of C.D. Broad published in 1925, was one of
the most discussed work. Even today a lot of researchers take this work as
the foundation for a philosophical definition of emergence. At the moment the
term emergence experienced a Renaissance in the philosophy of mind. We do not
want to go into further historical details here. The interested reader will
find a good treatment in \cite{Ste99}. Instead of this we want to present the
modern viewpoint of philosophy towards a definition of emergence.

Before we write down a philosophical definition of emergence we must define
what we mean by a system. The definitions we state here has to be understand in
a philosophical sense (so the formulations are very general) and it is a
different question if one can implement these definitions in a meaningful way
into natural sciences, like mathematics or physics. Furthermore it is the nature
of philosophical definitions that they contain fuzzy terms and concepts. We
cannot treat and discuss every detail here and refer for a more extensively
discussion to \cite{Ste99, Bec08}.

\begin{Def}
A \nt{system} consists of a set of components $C = \{C_1, C_2, \ldots \}$ and a
set of relations $R$ between these components. We denote a system by $S:=(C,R)$.
The properties of the components and of the relations are called
\nt{microstructure} of the system.
\end{Def}

\begin{Def}
A \nt{property} of a system is a characteristic feature of the system which is
reflected in the microstructure of the system. If a system has a property $E$,
but none of its components or subsets of components have the property $E$, then
we call this property a \nt{macro-property}.
\end{Def}

Based on the historical theories of emergence Stephan defines in \cite{Ste99}
different versions of emergence by stating characteristic features of systems
which have such emergent properties.

\begin{Def}[Weak emergence, \cite{Ste99}] \label{weakemergent}
A property $E$ of a system $S =(C,R)$ is called {\bf weak emergent}, if the
system has the following features.
\begin{itemize}
	\item [(i)] (physical monism) The system $S$ has only physical
components and every entity of the world is composed by physical components.
	\item [(ii)] (systemic property) The property $E$ is \nt{systemic}, that
means that no component or subset of components of the system have the property
$E$. Therefore $E$ is a macro-property.
	\item [(iii)] (synchrone determinacy or supervenience) The property $E$
depends nomologically on the microstructure of the system. The behaviour and
the properties of a system are therefore determined by the behaviour of its
components.
\end{itemize}
\end{Def}

The first item in the definition of a weak emergent property is a purely
philosophical requirement. In formal and mathematical theories this
requirement is out of debate since in natural sciences one always has the
belief that the world is assembled by physical components.
The second required feature for weak emergence is equivalent with that of a
macro-property, such as those occuring in statistical mechanics or other
theories.
The third required feature is a one-sided dependency relation, which said
that the macro-property depends on the microstructure of the system. That means
the following: The macro-property cannot change, if there is no change in the
microstructure of the system. There cannot exist another system with the same
macro-property but with a different microstructure. We say that the
macro-property \nt{supervenes} over the microstructure. Therefore in the
literature the term ``weak emergence'' is also known as supervenience. By
adding further features we can strengthen the term of weak emergence.

\begin{Def}[synchronous emergence, \cite{Ste99}]
A property $E$ of a system $S=(C,R)$ is called {\bf synchronous emergent}, if
$E$ is weak emergent and additionally has the following feature.
\begin{itemize}
	\item [(iv)] (irreducibility) The property $E$ is irreducible. That means
the property $E$ 
	\begin{itemize}
		\item [(a)] cannot be analyzed from the behaviour of the (isolated)
components of the system. This inability to determine that $S$ has property $E$
is a principle limitation (so even if we know everything one can know about the
single components it is still impossible to detect $E$ from that knowledge).
		\item [(b)] Or from the behaviour of the components of $S$ in different
constellations (with different relations) it is in principle impossible to
deduce that $S$ has property $E$.
	\end{itemize}
\end{itemize}
\end{Def}

The formulation that something is in principle impossible means that no
scientific progress can change that fact. Therefore synchronous emergence is
not an epistemological expression and not related to scientific knowledge
(often that crucial fact is misunderstood in the literature and emergence is
seen as an expression relative to scientific progress). 

There remains the question what exactly does it mean that a property $E$ can be
deduced from the microstructure of a system $S$. Broad does not say anything
about that in his work but Beckermann give an interpretation which shed some
light on it. He says the following: A property $E$ can be deduced from the
microstructure of a system $S$ if and only if one can deduce from the
{\em general laws of nature} that every system with that microstructure
consists of all features which characterises $E$ \cite{Bec08}.

So far we considered systems without a time component. Adding a time component
we can define an equivalent strong expression of emergence for time-depending
systems. In such systems the emergent property develops during time course.

\begin{Def}[diachrone emergence, \cite{Ste99}]
A property $E$ of a system $S = (C,R)$ is called {\bf diachrone emergent}, if
$E$ is weak emergent and additionally has the following features.
\begin{itemize}
	\item [(v)] (novelty) The property $E$ is genuinely new, that means that $E$
not appeared at an earlier time.
	\item [(vi)] (structure unpredictability) It is in principle impossible to
predict that property $E$ will appear during time course of the system.
\end{itemize}
\end{Def}

Like before the expression of diachrone emergence is not an epistemological
expression. Strictly speaking the novelty postulation means that a property $E$
of a system $S$ has never been seen before, even in other systems $E$ has not
be seen before. So $E$ appeared the first time ever. 

Unpredictability means that in principle one cannot predict a property $E$
of a system $S$ from the knowledge of the underlying microstructure of the
system $S$. Stephan argued in \cite{Ste99} that synchronous and diachrone
emergence are equivalent forms of emergence (up to the time component).

However in the context of stationary stochastic processes only synchronous
emergence is interesting, since a stochastic process with a diachrone emergent
property need to be non-stationary (see also the remarks in \cite{Set08}).

Beckermann defines synchronous emergence in a more compact way.

\begin{Def}[\cite{Bec08}] \label{BeckermannEmergent}
A macro-property $E$ of a system $S$ with microstructure $(C,R)$ is synchronous
emergent if and only if
\begin{itemize}
	\item [(a)] The sentence: ``All systems with microstructure $(C,R)$ own the
macro-property $E$'' is a valid law of nature, but
	\item [(b)] $E$ cannot be deduced in principle from the full knowledge of
all features the isolated components $C$ own or they have in different
arrangements.
\end{itemize}
\end{Def}

Postulation (a) is basically the same as supervenience (like in the definition
of weak emergence). Though (a) means a bit more. Beckermann stressed that the
sentence ``All systems with microstructure $(C,R)$ own the macro-property $E$''
is a valid law of nature has to be understood as follows: The law of nature is
not a special case of an already existing law of nature and cannot be deduced
by combining existing laws of nature. Consequently one has to discover this law
of nature for the first time and it has to be accepted as a law of nature. This
is a very strong requirement and one can see that this kind of (strong)
emergence appears very rarely. To be precise it is not clear at all if such a
strong version of emergence even exists in the real world. However the
definition of Beckermann and the definition of Stephan are equivalent. 

For our research and the treatment in this paper we take these philosophical
definitions of emergence as a basis. Because of the unclear situation depending
the existence of strong emergence in the real world we only consider weak emergence and try to
formalize this concept mathematically.

\subsection{Examples of Emergence} \label{BeispieleEmergenz}

Before we consider mathematical definitions of emergence we give a few
examples of emergent properties appearing in the real world.

\subsubsection{Weak Emergence}
There are numerous examples for weak emergence. We consider only three
well-known examples.

\begin{itemize}
	\item Flight structure of migratory birds and swarm behaviour in nature. \\
Observing swarms of animals (in particular swarms of birds or fishes) and their
behaviour is a fascinating spectacle. The swarm seems to have an own dynamic
which is not controlled by a central entity. Instead of this a kind of
self-organisation seems to be responsible for the dynamic. The behaviour of the
swarm supervenes over the single individuals. There are simple mathematical
models which model such a behaviour. For example Cucker and Smale showed
analytically for such a model (consisting of differential equations) that it
converges under certain preconditions against a stable solution \cite{Cuc07}.
In their work the dynamic of the centre of mass of the swarm is the emergent
macro-variable. The single trajectories of the swarm components corresponds to
the microstructure. Cucker and Smale showed that under some conditions the
centre of mass converge against a stable solution. That means that the whole
swarm behaviour developes from a chaotic looking behaviour to a well structured
behaviour. This well structured behaviour is the emergent property of the
dynamical system.

Instead of that Seth defined a measure for weak emergence, the so called {\em
G-emergence} (see \cite{Set08}). He calculates this measures for similar swarm
models. Changing different parameters in his model he observe numerically that
the G-emergence attains a higher value the more the swarm has a stable
movement structure. On the other hand if the individuals of a swarm behave
completely random the G-emergence attains values near zero.

	\item Neuronal networks. \\ 
A neural network (or artificial neural network) is a network imitated from the
network structure of neuronal cells in the human brain. It consists of neurons
and weighted connections between the neurons. The topology of the network is
usually fixed, so that the weights are the only changeable parameters. Every
neuron owns an appointed threshold and can accept input values from an external
user or from other neurons. This input is multiplied by the connection weight
and sumed up.\footnote{In general there are many different possibilities to
process the input values in a neuron. To simplify life we only consider one
possibility in this paper.} If this sum is higher than the threshold of the
neuron then the neuron {\it fires} an output signal to its successor neurons or
to the user. So the whole network works as follows: The user sends an input
signal into the network which is passed through the network and the user gets
back an output signal.

Neural networks are often used to classify objects or for forecast purposes.
For that the networks are initially trained with a labeled training set. To
minimize misclassifications one can change the weights between the neurons.
There are a lot of different learning algorithms, like the back propagation
algorithm, which are suitable for that task. After training the network
sufficiently well it can be used for new classification tasks. The big
advantage of a neural network is its flexibility and its ability to learn a
specific behaviour. From a mathematical point of view a neural network is a
dynamical system and one can show that under very mild assumptions it can
approximate every nonlinear and non continuous function. The disadvantage is
that it is a-priori not clear what kind of topology one has to choose to solve a
given classification problem with a neural network. We do not want to enter
closer into this problem and refer to \cite{Sta91} for a more detailed
treatment of that problem.

Instead of this we look at emergence in such networks. As a macro-property we
specify the classification ability of a trained network. The microstructure
consists of the neurons and the connections between them. It is obvious that
the macro-property is a systemic property, since no part of the microstructure
and no single neuron can have the ability to classify objects in the same way
as the whole network does. The macro-property also supervenes about the network
structure and the corresponding weights, because if one change some part of
the microstructure also the ability to classify objects will change. Because of
that the macro-property is a weak emergent property.

But the macro-property is a reducible property, since with the knowledge of the
microstructure one can completely explain (at least in theory) the
macro-property. Therefore the macro-property is not synchronous emergent (see
also Chapter 17 in \cite{Ste99}). Furthermore the learning process of the
macro-property is also not a case of diachrone emergence. One can calculate the
changes of the weights exactly and one can theoretically estimate when the
performance of a network is below a given error bound. The learning process
in a neural network is nothing else than an optimization of a multidimensional
function. So the macro-property is also not an example for a diachrone emergent
property.

	\item Phase transitions.\\ 
Everybody knows phase transitions from everyday life. For example consider the
change of fluid water to solid ice. This is considered as a phase transition.
In mathematical language a {\em phase} is defined as a pure probability measure
for a given model. Consider now the well-known Ising-model. One can show
that the set of asymptotic Gibbs measures is not empty and a convex set
\cite{Kna06}. A pure Gibbs measure is a Gibbs measure which cannot be written
as a convex combination of two other Gibbs measures. In the one dimensional
Ising-model there exists exactly one Gibbs measure and there is no phase
transition. In the two dimensional Ising-model there are two Gibbs measures in
the low temperature region and there is a phase transition at a critical
temperature. Below that critical temperature the system remains in one of the
two alternative states. The microstructure in the Ising-model consists of single
spins and the interactions between them, which are described by the energy
function. As a macro-property we can choose the {\em mean
magnetization}.\footnote{The mean magnetization is the mean value of the spin
values.} The mean magnetization is a systemic property by definition, since
every spin has a direction but does not reflect the characteristic features of the
mean magnetization, namely the disappearing variance. The mean magnetization
supervenes over the spins and the interactions between them. This is due to the
fact that if one changes the interactions between the spins than also the mean
magnetization will change. Therefore the spontaneous magnetization is a weak
emergent property.
\end{itemize}

\subsubsection{Strong emergence}
A rigoros proof for the existence of strong emergence in the real world as it
has been defined in the previous section is still missing. Some experts in the
theory of emergence say that the only serious example discovered so far for
strong emergence are mental states and similar phenomena of consciousness.
In philosophy, mental states are sensations like pain or intensions like
beliefs, hopes, etc. (\cite{Bec08}, p. 17). Such a mental state can be seen as a
macro-property of the human brain which is composed of physical
components considered as the microstructure. There is a wide acceptance among
the experts that a mental state like pain is determined by the underlying
microstructure and thus is a weak emergent property. But there are also a lot of
people who stress the fact that it is up to now impossible to reduce a
mental state to its physical microstructure (which are just physical states)
and thus explain it in a physical way. Some of the experts are convinced that
no progress in science can change that situation. There are also people who
claim the opposite. Another argument for the existence of strong emergence is
{\em downward causation}. This means that the direction of causality is
reversed. So the macro-property which was determined by the microstructure acts
now back to the microstructure and influence it. This kind of feedback loop
brings the whole system into a stable state. People who believe in downward
causation often give the following example. Suppose an individual has the mental
state fear. This mental state is determined by the underlying physical
structure, but one can measure an increase of the pulse and also the change of
lot of other physical properties can be measured (in this scenario the whole
physical body is the microstructure). So one can think that the mental state changes the physical structure and thus the microstructure of the system. Critics, however, are of the opinion that
a mental state cannot determine something. The problem is that a clear
definition of mental state is missing. In any case there is a big discussion
about that problem and for further readings we want the reader refer to
\cite{Ste99,Bec08, Cha02}.

There may certainly be a number of examples where one can suppose strong
emergence. For example Chalmers suppose that some phenomena appearig in quantum
physics could be considered as an example for strong emergence. But in his
treatment a clear argument is missing \cite{Cha02}. In a summary we can say
that up to now we are not sure if we can find strong emergence in the real
world and also no proof exists that show that we cannot find it.

\subsection{Mathematical Models for Emergence}

After defining and discussing the term emergence from a philosophical point of
view, we now want to look at it from a mathematical point of view. Indeed there
are some theories which have the ability to detect emergent phenomena but are
not able to give a clear definition how emergence can be understood in
mathematical terms. We just mention a small selection of the possible attempts
to formalize emergence. In particular we want to consider information theoretic
models for emergence. 

\subsubsection{Bifurcation Theory} \label{Bifurkationstheorie}

Bifurcation theory deals with the question if a solution of a
parametrized dynamical system is stable or not and with the question for which
parameters it becomes stable. For an introduction into the theory and a detailed
treatment see \cite{Guc83}. 

What is the connection between bifurcation theory and emergence? In this
section we try to give an answer to that question. Consider a
parametrized dynamical system which is described by a set of equations (for
example a system of differential equations). These equations describe in an
implicit way the microstructure of the system (implicit because typically only
macro variables appear in the equations). This microstructure can be changed
via changing the parameters. One can imagine the single components of the
system as solution curves of the dynamical system for different initial values.
As a macro-property one can choose multistability of the system (that means
that there exists several stable solution branches). This macro-property is a
systemic property, since no single solution can have the property of
multistability. Furthermore the multistability depends directly on the
microstructure, since it changes with changing some parameters (remember that
parameters belong to the microstructure) of the system. So multistability is a
weak emergent property. Since one can (at least in principle) determine from
the equations for which parameter values a bifurcation occurs and, thus
multistability, the property of multistability is not strong emergent and can
be deduced from the microstructure.
In summary we can say that with bifurcation theory one can detect cases of weak
emergence, but it is not clear if every weak emergent property can be detected
in that way. There are situation in non-equilibrium in which it is difficult to
detect bifurcations. Maybe there are also much more complicated emergent
properties in nature that cannot be modeled by such kind of systems. Furthermore
a direct link to the microstructure of the system is missing and that is a
further reason why we follow an information theoretic approach for defining
emergence.

\subsubsection{Synergetics} \label{Synergetik}

Another model which is related to bifurcation theory is synergetics which was
introduced by Haken in the 1960s \cite{Hak83}. Haken tried with his
theory to explain the evolution of new system properties. Often he considered
structures which appear spontaneously through a self-organisation process. A
classical example is the appearance of laser light from an ordinary light
source which is feed permanently with energy from outside. After exceeding a
certain amount of energy laser light appears. From a mathematical point of
view one can consider synergetics as a method to approximate solutions of high
dimensional differential equations, see \cite{Jet89, Hak83} for examples.
Basically Haken introduce few artificial macro variables which determine the
main behaviour of a system of equations and neglect the other remaining
variables. So one can imagine that these macro variables determine the
behaviour of the microstructure (that would be a case of downward causation).
But it is not clear if the macro variable can be deduced from the
microstructure, since it was introduced artificially. Due to this fact it is
not even clear if such macro variables can be considered as properties of the
system. At least from Haken's point of view it remains questionable if such
macro variables can be seen as emergent or not.

\subsection{Information Theoretic Definitions of Emergence}

Next we want to have a look at some information theoretic approaches to
formalize emergence. We will shortly discuss two different approaches which are
related to excess entropy and statistical complexity. In particular we discuss
if complexity measures like the $PMI$ are suitable to detect emergence.

\subsubsection{Emergence as Reduction of Complexity} \label{ShaliziEmergenz}

Shalizi and Crutchfield \cite{Sha01,Cru94} suggest a mathematical definition of
emergence as follows. First they define a quantity which measure the efficiency
of prediction of a stochastic process.

\begin{Def}[Efficiency of prediction, \cite{Sha01}]
The \nt{efficiency of prediction} of a stationary stochastic process is the
ratio between its excess entropy and its statistical complexity
$$e^+ := \frac{E}{C_P^+}, \qquad e^- := \frac{E}{C_P^-},$$
where $e^+:= 0$ if $C_P^+ = 0$ and $e^- := 0$ if $C_P^- = 0$.
\end{Def}

From the properties of the excess entropy it follows that
$$
0 \le e^+ \le 1, \qquad 0 \le e^- \le 1.
$$
The efficiency of prediction tells us how much of the internal process
information can be actually used for predicting future process behaviour.

\begin{Def}[Derived process, \cite{Sha01}] \label{derivedProcess}
A stationary stochastic process \\
$(S_t', t \in \setZ)$ is called {\bf derived} from another stationary stochastic
process $(S_t, t \in \setZ)$ if and only if there is a measurable function 
$f: \overleftrightarrow{\mathbf{S}} \rightarrow G$ in a measure space $(G, \kG)$
such that $S'_t := f(\overleftarrow{S}_t)$. $(S_t', t \in \setZ)$ is called
the derived or filtered process and the function $f$ is denoted as filter.
\end{Def}

Based on that Shalizi defines emergence as follows.

\begin{Def}[Emergent Process, \cite{Sha01}] \label{ShaliziEmergent}
A derived stochastic process is emergent, if it has a greater efficiency of
prediction $e^+$ than the process it derives from. We then say the
derived process emerges from the underlying process.
\end{Def}

\begin{Def}[Intrinsic Emergence, \cite{Sha01}]
A process is \nt{intrinsic emergent}, if there is another process which
emerges from it.
\end{Def}

Shalizi justified his definition on the following basis. At the one hand
Shalizi's idea of emergence is that an emergent property supervenes over the
components of a system (this idea coincide with the definition of weak
emergence seen before). On the other hand he assumes that the appearance of
emergence implies a simplified description of the system. This idea he
describes with a reduced complexity as it is formalized in Definition
\ref{ShaliziEmergent}. Shalizi says that Definition \ref{derivedProcess}
represents the assumption of supervenience. It remains questionable if based on
such a vague argument a reasonable definition of emergence is possible. The
problem is that Shalizi not clearly defines what he means with system and
emergent properties in a philosophical sense. Because of that lack of
philosophical basis we use the definitions of emergence from Section
\ref{PhilEmergentDef}. 

Within this setting we had to consider a derived process
$\overleftrightarrow{S}'$ as a macro-property of the underlying
process $\overleftrightarrow{S}$ and the random variables $S_t$ together with
their correlations are forming the microstructure. By definition the
macro-property is a systemic property. Furthermore $\overleftrightarrow{S}'$
supervenes over the microstructure, since if one changes the underlying process 
$\overleftrightarrow{S}$ in general the process $\overleftrightarrow{S}'$ will
also change (but there exists filters such that this is actually not the case).
In such situations the derived process can be classified as weak emergent. If
the derived process additionally has a higher efficiency of prediction than the
underlying process, than there is more information stored in the realisations of
the derived process as in the underlying process. In the extreme case one has
$C_P^+ = E$ and with Corollary \ref{statCompExzessEnt} that is the case if
and only if $S^+ = g(S^-)$. That means that the future causal states can be
completely deduced from the past causal states. 

In the derived process we can ``better'' deduce future causal states from past
causal states as in the underlying process but it is not clear how this is
related to strong emergence.

In \cite{Sha01} Shalizi also state a concrete filter function to construct a
derived process. Unfortunately further examples and results are missing and
more evidence (in form of examples) are necessary to check if this definition
of emergence is reasonable or not. But nevertheless it is an interesting
approach.

\subsubsection{Model of Emergent Description}

In \cite{Pol04, Pol06} Polani suggest an emergent description of dynamical
systems. Inspired by the theory of synergetics by Haken, he states an
information theoretic decomposition of a dynamical system into
information-preserving and independent subsystems. We consider the set of all
realisations $\overleftrightarrow{\mathbf{S}}$ of a stationary stochastic
process.
Polani decomposes this set into finite many components. Such a decomposition is
given in form of $k$ random variables $\overleftrightarrow{S}^{(i)} :
\overleftrightarrow{\mathbf{S}} \rightarrow \kS^{(i)}$, $i = 1, \ldots, k$ with
$\bigcup_{i =1}^k \kS^{(i)} = \overleftrightarrow{\mathbf{S}}$. The random
variables are not further specified and it remains unclear under which
conditions such a decomposition exists. If one assumes that it exists then one
can imagine it like depicted in the following diagram.

\xymatrixcolsep{2pc}
\xymatrixrowsep{2pc}
$$\xymatrix{ \Omega \ar[d]_{\overleftrightarrow{S}} &  & \kS^{(1)} \\
\overleftrightarrow{\mathbf{S}} \ar[rr]^{\overleftrightarrow{S}^{(2)}} \ar[rru]^{\overleftrightarrow{S}^{(1)}} \ar[rrd]^{\overleftrightarrow{S}^{(k)}} & & \kS^{(2)} \ar@{.}[d] \\
& & \kS^{(k)} }$$

With that Polani defines the emergent description of a process.

\begin{Def}[Emergent description, \cite{Pol06}]
Let a stationary stochastic process with a decomposition in $k$ random
variables be given. Then the $k$ random variables
$(\overleftrightarrow{S}^{(1)}, \ldots, \overleftrightarrow{S}^{(k)})$ are
called an {\bf emergent description} of $\overleftrightarrow{S}$, if 
\begin{itemize}
	\item [(a)] the decomposition is a complete representation of the systems:
		$$I(S_t ; S_t^{(1)}, \ldots, S_t^{(k)}) = H(S_t),$$
	\item [(b)] the individual components of the decomposition are independent
					of each other:
		$$I(S_t^{(i)} ; S_t^{(j)} ) = 0 \quad \forall \; i \neq j,$$
	\item [(c)] and the components are information-preserving in time:
		$$I(S_t^{(i)} ; S_{t+1}^{(i)} ) = H(S_{t+1}^{(i)}).$$
\end{itemize}
\end{Def}

\begin{figure}[h!]
\begin{center}
\epsfig{figure=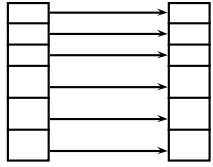,scale=0.4}
\caption{Schematic scetch of an emergent description as a decomposition
into independent components.}
\phantomsection \label{fig:emergentDescription}
\end{center}
\end{figure}

The significant difference to the previous approaches is that the whole
realisation space will be decomposed and thus the whole process will be
decomposed. Figure \ref{fig:emergentDescription} shows a schematic scetch of
that situation.

An emergent description is a decomposition in information-independent
components which preserve their information for all times. Unfortunately also
for that model there are no results known which guarantee existence of such a
decomposition. Even Polani does not give any explicit example in his work.
Because of that lack of knowledge it is difficult to judge this proposal in a
reasonable way. With the facts known up to now it seems not possible to check
the relation of that description to the definition of emergence given in
Section \ref{PhilEmergentDef}. 

\subsubsection{Complexity Measures as Definitions of Emergence}

We now want investigate how well the complexity measures excess entropy,
statistical complexity and persistent mutual information suit to define and
detect emergence. We shortly repeat how one can understand and interpret the
different complexity measures.
\begin{itemize}
	\item Excess entropy $E$: The amount of past information which is currently
available and communicated into the future. In particular it represents the
amount of information one can extract from a concrete (past) realisation to make
predictions for future realisations.
	\item Statistical complexity $C_P^+$: Amount of information stored in the
future causal states. In other words it is the amount of information about
the past which is stored in the process to predict future in an optimal way. In
general a concrete realisation contains less information than the process has
internally stored.
	\item Persistent mutual information $PMI$: The amount of past information
which is currently available and which is communicated into a very far future.
In other words: the amount of information one gets from a concrete
past realisation and which will be preserved for all future times and for all
future realisations.
\end{itemize}

As we proved before we have
$$PMI \le E \le C_P^+.$$

So we have a gradiation from a fine complexity measure $C_P^+$ to a coarse one,
the $PMI$. If we consider regularly structures, like $L$-periodic processes,
then we establish that all three complexity measures coincide.

Such processes can be generated by dynamical systems with a periodic behaviour.
One example is the logistic map which produce periodicity for certain parameter
values. It is defined with a parameter $r$ as follows
$$
T_r : [0,1] \rightarrow [0,1], \qquad T_r (x) := rx(1-x), \qquad r \in [0,4],
$$
$$
x_{n+1} := T_r(x_n), \qquad x_0 \in [0,1].
$$
Furthermore we define a random variable $S: [0,1] \rightarrow \kA$ by
$$
S(x_n) := \left \{ \begin{array}{l l} 0, & x_n \in [0, 0.5], \\ 1, & x_n \in
(0.5, 1], \end{array} \right.
$$
and the alphabet $\kA := \{0,1\}$. The discrete time series which is produced
by 
$x_{n+1} = T_r(x_n)$ for an arbitrary initial value $x_0 \in [0,1]$, is called
{\bf trajectory}. Together with a $\sigma$-algebra and a $T$-invariant
probability measure\footnote{The measure depends on $r$ and its existence is a
priori not clear.} we get a stationary stochastic process. A subset $A$ of the
phase space $[0,1]$ is called invariant under $T_r$, if $T_r(A) \subset A$.

\begin{Def}
A closed invariant set $A$ is called {\bf attracting set}, if there is an
environment $U$ of $A$ such that for the flow $(T_t)_{t \in \setR}$ of $T$ it
holds that
$$
\lim_{t\rightarrow \infty} d(T_t(x) , A) = 0, \qquad \forall \; x \in U.
$$
$d$ denotes the corresponding metric in the phase space.
\end{Def}

\begin{Def}
A (compact) invariant set $A$ of the phase space is called {\bf attractor}, if
$A$ is an attracting set which contains a dense trajectory.
\end{Def}

For suitable initial values all trajectories of a dynamical system tend to an
attractor of the dynamical system.

\begin{figure}[h]
\begin{center}
\epsfig{figure=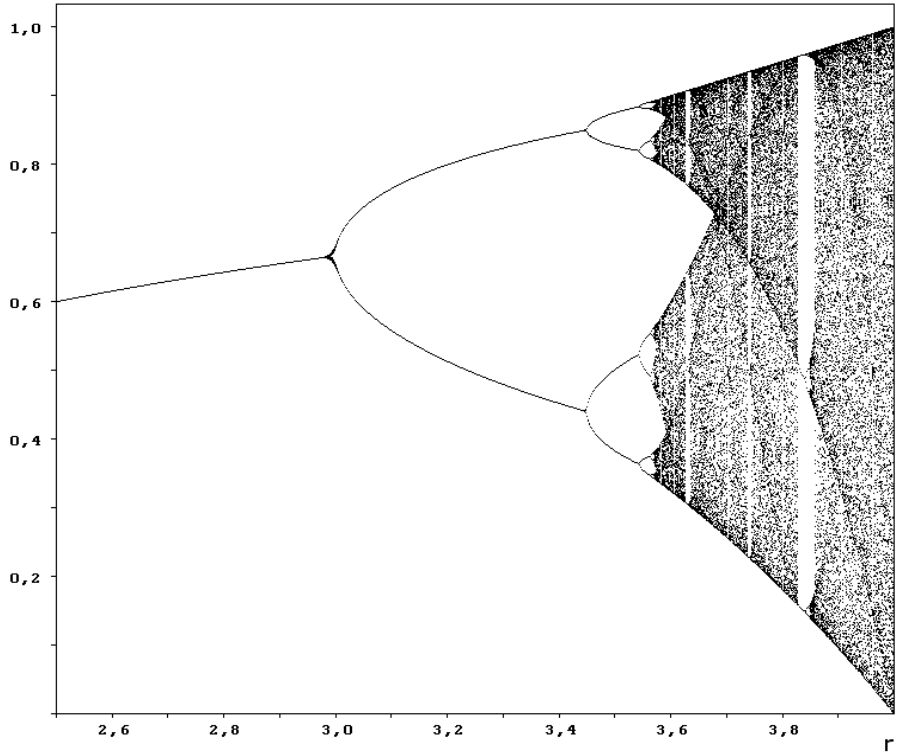,scale=0.8}
\caption{Attractor of the logistic map for parameter values $r \in [2.5,4]$.}
\phantomsection \label{fig:logAttractor}
\end{center}
\end{figure}

We can draw an attractor in a graphical way to get an overview of the
long-term behaviour of the dynamic.\footnote{For that we fix $r$ and choose an
initial value $x_0$ and calculate the corresponding trajectory. Now one draws
the points of the trajectory depending on the parameter starting after a few
hundred iterations to avoid numerical artifacts.} Figure
\ref{fig:logAttractor} shows the attractor of the logistic map for parameters 
in the intervall $[2.5, 4]$.

In the lower parameter region one observe a periodic behaviour of the logistic
map. For example if we pick $r=3.2$ then the period is $2$. The random variable
$S$ codes such a period in the generated symbol sequence such that we get a
periodic stochastic process. For such processes the values of $C_P^+, E$ and
$PMI$ are equal. If we look at the attractor we recognize some parameter values
for which the period doubles if we pass them. At this points a bifurcation
occurs and new solution branches appear. The periodic behaviour corresponds to
multistability we considered in Section \ref{Bifurkationstheorie}. We have
already seen that multistability is a weak emergent property of the system. All
three complexity measures detect this weak emergent property due to the non
trivial values they have. Up to that position every complexity measure is equal
powerful to detect weak emergent properties. But when we ask if every structure
is weak emergent where these complexity measures are assuming positive values,
we will see the differences. Table \ref{tab:Beispiele} shows a summary of
different examples together with some calculated entropic expressions
and complexity measures.
\\

\begin{center}
\begin{tabularx}{15.0cm}{|C{1.5cm}||C{1.8cm}|C{2.7cm}|C{1.2cm}|C{1.2cm}|C{1.9cm}
|C{1.6cm}|}
  \hline
  Feature & $p$-periodic & $R$-Markov & i.i.d. & Thue-Morse & $1$-D Ising
model & perfect random \\
  \hline\hline
  $h_P$ & $0$ & $\triangle H(R+1)$ & $H(1)$ & $0$ & $*$\footnote{$\log\lambda_1
- \frac{\beta}{\lambda_1} \frac{d}{d\beta}\lambda_1$} & $\log|\kA|$ \\ \hline
  $E$ & $H(p)$ & $H(R)-Rh_P$ & $0$ & $\infty$ & $H(1)-h_P$ & $0$ \\ \hline
  $C_P^+$ & $H(p)$ & $H(R)$ & $0$ & $\infty$ & $H(1)$ & $0$ \\ \hline
  $C_P^-$ & $H(p)$ & $H(R)$ & $0$ & $\infty$ & $H(1)$ & $0$ \\ \hline
  $PMI$ & $H(p)$ & $0$ & $0$ & $\infty$ & $0$ & $0$ \\ \hline
  $e$ & $1$ & $1-\frac{R}{H(R)}h_P$ & $0$ & ? & $1-\frac{h_P}{H(1)}$ & $0$ \\
\hline
\end{tabularx}
\end{center}
\begin{table}[h]
\caption{Tabular overview of structure quantities corresponding to example
processes.}
\phantomsection \label{tab:Beispiele}
\end{table}

Consider the one-dimensional Ising-model. In the low temperature region the
excess entropy assumes positive values (compare Figure \ref{fig.excessIsing}).
Also the statistical complexity give positive values. We already know that
there is no phase transition in the one-dimensional Ising model. So there is no
weak emergent property to detect (at least if we choose the philosophical
definition given in the section before). This indicates that the excess entropy
and the statistical complexity are not suitable for detecting weak emergent,
since they detect something (assume positive values) although there is nothing
interesting to detect (in the sense of emergence). It seems that $E$ and
$C_P^+$ can detect interesting structures of dynamical systems but they are too
fine to filter weak emergent structures. Only the persistent mutual information
is zero in that examples (except for the Thue-Morse example). So the only
remaining candidate for detecting emergent properties is the $PMI$. But the
calculated examples so far are not enough to decide if $PMI$ really detect
emergent properties. In particular an example is still missing where $PMI$
assumes a non trivial value and clearly differs from $E$ and $C_P^+$. But there
is some numerical evidence. Ball, Diakonova and MacKay numerically calculated in
\cite{Bal10} the $PMI$ for the logistic map. From the numerical results one
recognize that in the parameter region $r \in [3.58, 3.68]$ so called ``chaotic
bands'' appear which are detected by the $PMI$. Roughly speaking chaotic bands
are disjoint regions in the phase space $[0,1]$ between which a trajectory
changes periodically but inside the region behaves in a chaotic way. If this is
a further example for a weak emergence property is an open question. In
particular a rigoros analytical investigation and calculation of the $PMI$ for
that parameter values is missing. 

In summary we can say that the $PMI$ is the most promising complexity measure
among the ones investigated here for detecting weak emergent properties of
systems. But there are still examples and rigoros results missing to
further confirm this conjecture.

\subsection*{Acknowledgment}
I would like to thank Andreas Knauf for motivating me to work on this
project, for his constant support and many useful discussions and suggestions.

\begin{appendix}

\section{Information-theoretic Facts}

\begin{Satz} \label{InfoProperties}
For the mutual information and two random variables $X,Y$ it holds
\begin{itemize}
	\item [(i)] $I(X;Y) \ge 0,$ with equality iff $X$ and $Y$ are stochastic indpendent.
	\item [(ii)] $I(X;Y) = I(Y;X).$
	\item [(iii)] Let $X= (X_1, X_2, \ldots)$ be a stochastic process then it
			holds that
		$$I(X;Y) = \lim_{n \rightarrow \infty} I((X_1, \ldots, X_n); Y).$$
\end{itemize}
\end{Satz}
\begin{Proof}
See \cite{Pin64} Chapter 2.2.
\end{Proof}

\section{Excess Entropy}

The metric entropy is defined by $h_P := \lim_{L \rightarrow \infty}
\frac{H(L)}{L}$ and give us a geometric interpretation of the excess entropy.
\begin{Satz}[\cite{Gra86}] \label{SatzConv}
It holds that
$$E = \lim_{L \rightarrow \infty} (H(L) - L h_P),$$
\end{Satz}
\begin{Proof}
We write $E$ as the limit of partial sums and use discrete integration 
$$
\sum_{L=1}^M (\triangle H(L) - h_P) = H(M) - H(0) - M h_P .
$$
Because of $H(0) = 0$ it follows that
$$
E = \lim_{M \rightarrow \infty} (H(M) - M h_P ).
$$
\end{Proof}

\begin{Satz}[\cite{Ell09}] \label{ExcessEntKausal}
For a stationary stochastic process $\overleftrightarrow{S}$ it holds that
$$E = I(\kS^+; \kS^-).$$
\end{Satz}
\begin{Proof}
To prove the proposition we use a four random variable mutual information
introduced in \cite{Yeu91} and follow the same strategy as in \cite{Cru10}. For
random variables $X, Y, Z, U$ we define
\begin{eqnarray*}
I(X; Y; Z; U) & := & I(X;Y;Z) - I(X; Y; Z |U), \\
I(X;Y;Z) & := & I(X;Y) - I(X;Y|Z), \qquad \; \; \; \\
& & {\rm with \;} I(X;Y|Z) := H(X|Z) - H(X|Y,Z), \\
I(X; Y; Z |U) & := & I(X;Y|U) - I(X;Y|Z;U), \; \; \\
& & {\rm with \;} I(X;Y|Z;U) := H(X|Z,U) - H(X |Z,U,Y).
\end{eqnarray*}
Furthermore we use the following two identities which hold for a
measurable function $f$ of a random variable $X$ (see Lemma 2.5.2
in \cite{Gra90})
\begin{equation} \label{ProofhelpEquations}
H(f(X)|X) = 0, \qquad H(X, f(X)) = H(X).
\end{equation}
We define mappings $g:\kA^\setN \rightarrow \{1, \ldots ,m\}$ with $g(\sigma) :=
j$ if $\sigma \in S_j^-$ and $f:\kA^{-\setN_0} \rightarrow \{1, \ldots,
n\}\times \kA$ with $f(\sigma \sigma_0) := (i,\sigma_0)$ if $\sigma \in S_i^+$.
Since we are considering $\epsilon$-machines the mappings $g$ and $f$ are well-defined and measurable. Thus we can write $\kS^+
=f(\overleftarrow{S}), \kS^-=g(\overrightarrow{S})$ and using
(\ref{ProofhelpEquations}) we get
\begin{eqnarray}
H(\kS^+ | \overleftarrow{S}) & = & 0,  \qquad \qquad \qquad \, \,H(\kS^- |
\overrightarrow{S}) = 0,  \label{condEntrFuncDisappear} \\
H(\overleftarrow{S}, \kS^+ ) & = & H(\overleftarrow{S}),   \qquad \qquad
H(\overrightarrow{S}, \kS^-) = H(\overrightarrow{S}), \\ 
H(\overrightarrow{S}| \overleftarrow{S}, \kS^+ ) & = & H(\overrightarrow{S}
|\kS^+),
  \qquad \; H(\overleftarrow{S} | \overrightarrow{S}, \kS^-) =
H(\overleftarrow{S} | \kS^-). \label{condEntropyDisappear}
\end{eqnarray}
In the next step we show $I(\overrightarrow{S};\overleftarrow{S}; \kS^+; \kS^-) =
I(\overrightarrow{S}; \overleftarrow{S}) = E$.
Consider
\begin{equation} \label{secondfourMITerm}
I(\overrightarrow{S}; \overleftarrow{S} ; \kS^- | \kS^+) = I(\overrightarrow{S}
;\overleftarrow{S} | \kS^+) - I(\overrightarrow{S} ; \overleftarrow{S} | \kS^+ ;
\kS^-),
\end{equation}
and remark that the first term vanishes because with
(\ref{condEntropyDisappear}) it holds that
$$
I(\overrightarrow{S} ; \overleftarrow{S} | \kS^+) = H(\overrightarrow{S} |\kS^+)
-H(\overrightarrow{S} | \overleftarrow{S} ,
\kS^+) \stackrel{\left(\ref{condEntropyDisappear}\right)}{=} 0.
$$
The second expression of (\ref{secondfourMITerm}) is also zero, since
$$
I(\overrightarrow{S}; \overleftarrow{S} | \kS^+ ; \kS^-) =
H(\overrightarrow{S}|\kS^+ , \kS^-) - H(\overrightarrow{S} | \kS^+ , \kS^- ,
\overleftarrow{S}) \stackrel{\left (\ref{condEntropyDisappear}\right)}{=} 0.
$$
Putting all together we yield
$$
I(\overrightarrow{S}; \overleftarrow{S} ; \kS^- | \kS^+) = 0.
$$
Furthermore we have
$$
I(\overrightarrow{S}; \overleftarrow{S}; \kS^-) = I(\overrightarrow{S};
\overleftarrow{S}) - I(\overrightarrow{S}; \overleftarrow{S} |\kS^-) =
I(\overrightarrow{S}; \overleftarrow{S}),
$$
since $I(\overrightarrow{S} ; \overleftarrow{S} |\kS^-) = H(\overleftarrow{S} |
\kS^-) - H(\overleftarrow{S} | \overrightarrow{S} , \kS^-)
\stackrel{\left(\ref{condEntropyDisappear}\right)}{=} 0$.
Putting things together we get
$$
I(\overrightarrow{S}; \overleftarrow{S}; \kS^+; \kS^-) = I(\overrightarrow{S};
\overleftarrow{S}).
$$
In a second step we show $I(\overrightarrow{S}; \overleftarrow{S}; \kS^+;
\kS^-) = I(\kS^+; \kS^-)$. As in the first step the following term vanish
\begin{equation} \label{secondfourMITerm2}
I(\kS^+; \kS^- ; \overrightarrow{S} | \overleftarrow{S}) = I(\kS^+ ; \kS^- |
\overleftarrow{S}) - I(\kS^+; \kS^- | \overrightarrow{S}; \overleftarrow{S}) =
0,
\end{equation}
since $I(\kS^+; \kS^- |\overleftarrow{S} ) = H(\kS^+|\overleftarrow{S}) -
H(\kS^+| \kS^- ,\overleftarrow{S})
\stackrel{\left(\ref{condEntrFuncDisappear}\right)}{=} 0$ and
$$
I(\kS^+; \kS^- | \overrightarrow{S}; \overleftarrow{S}) = H(\kS^+ |
\overrightarrow{S}, \overleftarrow{S}) - H(\kS^+ | \kS^- , \overrightarrow{S},
\overleftarrow{S}) \stackrel{\left(\ref{condEntrFuncDisappear}\right)}{=} 0.
$$
We consider now
$$
I(\kS^+; \kS^-; \overrightarrow{S}) = I(\kS^+; \kS^-) -I(\kS^+;
\kS^-|\overrightarrow{S}),
$$
and the second term disappear, since
$$
I(\kS^+; \kS^- | \overrightarrow{S}) = H(\kS^- | \overrightarrow{S}) - H(\kS^- |
\kS^+ ,\overrightarrow{S})
\stackrel{\left(\ref{condEntrFuncDisappear}\right)}{=} 0.
$$
This results in 
$$
I(\overrightarrow{S}; \overleftarrow{S}; \kS^+; \kS^-) = I(\kS^+; \kS^-),
$$
and finally we get
$$
E = I(\overrightarrow{S}; \overleftarrow{S} ) = I(\kS^+; \kS^-).
$$
\end{Proof}

\begin{Korollar}[\cite{Ell09}] \label{statCompExzessEnt}
It holds that
$$C_P^+ = E + H(\kS^+ | \kS^-),$$
$$C_P^- = E + H(\kS^- | \kS^+).$$
Furthermore the following inequalities hold
$$C_P^+ \ge H(\kS^+ | \kS^-), \qquad C_P^- \ge H(\kS^- | \kS^+).$$
\end{Korollar}
\begin{Proof}
The first two claims follow with $E = I(\kS^+ ;\kS^-) = H(\kS^+) - H(\kS^+ |
\kS^-)$, $C_P^+ = H(\kS^+)$ and the symmetry of mutual information. Due to $E
\ge 0$ the other two inequalities follows.
\end{Proof}


\section{Spectral Analysis of Substitution Systems} \label{AppendixSpectral}

This section is an excerpt of Chapter $5$ in \cite{Que87}. We only state the
proofs which are relevant for us and refer for the remaining parts to \cite{Que87}. In
the following we consider a special type of dynamical systems, which often leads
to interesting sequences of symbols. \\
As usual we denote with $\kA$ a finite alphabet, e.g. $\kA := \{0, \dots, s-1\}$.
Furthermore we define $\kA^* := \displaystyle \cup_{k \ge 1} \kA^k$ as the
set of all finite words over $\kA$.
\begin{Def}
A mapping $\zeta : \kA \rightarrow \kA^*$ is called a \nt{substitution} on
$\kA$. To every letter $i \in \kA$ we assign a word $\zeta(i)$ such that for at
least one letter $i$ it holds that $l_i := |\zeta(i)| \ge 2$. If $l_i = q \ge 2$
holds for all $i \in \kA$, then $\zeta$ is a substitution of constant length
$q$.
\end{Def}

Every substitution $\zeta$ induces a mapping $\zeta : \kA^* \rightarrow \kA^*$
with
$$
\zeta(B) := \zeta(b_0) \ldots \zeta(b_n), \qquad B= b_0 \ldots b_n \in \kA^*.
$$

Similar we define a mapping $\zeta : \kA^\setN \rightarrow \kA^\setN$. We equip 
$\kA^\setN$ with the discrete topology such that the mapping $\zeta$ is
continuous with respect to this topology. Remark that in general $\zeta$ is not
surjective. $\zeta^k$ denotes the $k$-th iterative of $\zeta$. Fixed points of
$\zeta^k$ for a $k \ge 1$ are of special interest for us and the next
proposition give sufficient conditions for the existence of a fixed point.
\begin{Satz} \label{ExistenceFixpoint}
Let $\zeta$ be a substitution with $|\zeta^n(\alpha)| \stackrel{n \rightarrow
\infty}{\rightarrow} \infty$ for every $\alpha \in \kA$. Then there exists a
fixed point $u \in \kA^\setN$ and an integer $k \ge 1$, such that $u =
\zeta^k(u)$.
\end{Satz}
Henceforth we assume that $\zeta$ fulfills the following two conditions
\begin{equation} \label{FirstCond}
\displaystyle \lim_{n \rightarrow \infty} | \zeta^n(\beta)| = \infty, \qquad
{\rm for \; all} \; \beta \in \kA, 
\end{equation}
\begin{equation} \label{SecondCond}
{\rm there \; exist} \; \alpha \in \kA \; {\rm (denoted \; as \; 0 \;
in \; the \; following), \; such \; that \; the \; word} \; \zeta(\alpha)
{\rm \; starts \; with} \; \alpha.
\end{equation}

In particular this condition guarantees the existence of a fixed point, which
we denote as $u$ in the following. From now on the alphabet $\kA$ consists only
of those letters which actually appear in the word $\zeta^n(0)$ for all $n \ge
0$. As an example we consider the substitution which generates the Thue-Morse
sequence. With $\kA = \{0,1\}$ and $\zeta(0) = 01$, $\zeta(1) = 10$
the Thue-Morse sequence is the fixed point $u = \zeta^\infty (0)$. \\
For $\zeta$ and $u$ we associate a topological dynamical system $(X,T)$, with
$T: \kA^\setN \rightarrow \kA^\setN$ as the one-sided shift mapping on
$\kA^\setN$ and $X := \overline{O(u)}$ where $O(u) := \{T^n (u) : n \ge 0\}$.

We want to introduce the concept of ergodicity for the associated system and
define for that the notion of minimality.

\begin{Def}
A topological dynamical system$(X,T)$ is called \nt{minimal}, if the
$T$-invariant sets in $X$ are only $X$ and $\emptyset$.
\end{Def}

Minimality is characterized as follows.
\begin{Satz}
The system $(X,T)$ is minimal $\iff$ $O(x)$ is dense in $X$ for every $x \in X$.
\end{Satz}

In particular for the associated system the following result holds.
\begin{Satz} \label{MinCondition}
The system $(X,T)$ is minimal if and only if for every
$\alpha \in \kA$ there exist an integer $k\ge 0$, such that $\zeta^k(\alpha)$
contains $0$.
\end{Satz}

\begin{Def}
A substitution $\zeta$ is called \nt{irreducible} on $\kA$, if for every pair
of letters $\alpha, \beta \in \kA$ an integer $k = k(\alpha, \beta)$ exist, such
that $\beta \in \zeta^k(\alpha)$. $\zeta$ is called \nt{primitive}, if there
exist an integer $k$ independent of $\alpha,\beta$, such that $\beta \in
\zeta^k(\alpha)$ for all $\alpha, \beta \in \kA$.
\end{Def}

The condition in Proposition \ref{MinCondition}, which guarantees minimality,
implies that that $\zeta$ is primitive. If $\zeta$ is primitive then $X$ is not
depending on the fixed point $u$ instead it only depends on $\zeta$, since
every letter appears in $u$. Because of this the system $(X,T)$ is uniquely
determined through $\zeta$ and we denote it sometimes as $(X(\zeta),T)$.

For two words $B,C \in \kA^*$ we denote with $L_C (B)$ the number how often the
word $C$ appears in $B$. In particular for a letter $i \in\kA$ we write $L_i(B)$
for the number the letter $i$ appears in $B$.

\begin{Def}
The $s \times s$-matrix $M = M(\zeta)$ with $m_{ij} = L_i(\zeta(j))$ for $i,j
\in \kA$ is called \nt{$\zeta$-matrix}.
\end{Def}
$M$ is a positive $s \times s$-matrix with nonnegative integer entries. For
every $j \in \kA$ we have $\langle L(\zeta (j)), \setId\rangle = \displaystyle
\sum_{i \in\kA} L_i (\zeta(j)) = | \zeta(j)|$, where $\langle .,. \rangle$ is
the scalarproduct in $\setR^s$. For a word $B \in \kA^*$ we denote with $L(B)$ a vector in
$\setR^s$ with entries $L_i(B)$ for $0 \le i \le s-1$. It holds that
$L(\zeta(B)) = M \cdot L(B)$ and in particular $L(\zeta(j)) = (m_{ij})_{i \in
\kA}$. We denote $L : \kA^* \rightarrow \setR^s$ as \nt{composition-function}
and $M$ also as \nt{composition-matrix}. Remark that $\zeta$ is primitive if
$M(\zeta)$ is primitive, i.e. $M^k$ has positive entries for a $k$. The next
proposition gives interesting properties about primitive matrices, which
are crucial in the following treatment.

\begin{Satz}[Perron-Frobenius]
Let $M$ be a primitive, positive matrix. Then it holds that
\begin{itemize}
	\item [(a)] $M$ has a strictly positive eigenvalue $\Theta$, such that
$\Theta > |\lambda|$ for all eigenvalues $\lambda$ of $M$ which are different
from $\Theta$.
	\item [(b)] There exist a strictly positive eigenvector for $\Theta$.
	\item [(c)] $\Theta$ is a simple eigenvalue.
\end{itemize}
\end{Satz}
The dominating eigenvalue $\Theta$ is also called Perron-Frobenius eigenvalue
(PF-eigenvalue). A positive matrix is called \nt{irreducible}, if for
every $i,j$ an integer $k\ge 1$ exist, such that $m_{ij}^{(k)} > 0$.

\begin{Bem}
With the weaker assumption of an irreducible matrix one can almost show the
result of Perrron-Frobenius analogously. Only part $(a)$ changes as follows.
$M$ has a strictly positive eigenvalue such that $\Theta \ge |\lambda|$ for
every eigenvalue $\lambda$ of $M$ different from $\Theta$. We can
classify the eigenvalues with $\lambda \neq \Theta$ and $|\lambda | = \Theta$
with the help of periodicity exactly. 
\end{Bem}

\begin{Def}
The \nt{period} $d \ge 1$ of an irreducible, positive matrix $M$ is the
smallest common divisor of the set $\{k \ge 1: m_{ii}^{(k)} > 0\}$ for every
$i$.
\end{Def}

In particular we have the following relation.
\begin{Satz}
An irreducible, positive matrix is primitive if, and only if the period is
$d=1$.
\end{Satz}

\begin{Satz}
Let $M$ be an irreducible, positive matrix with period $d>1$, then there are
exactly $d$ eigenvalues $\lambda$ of $M$ with $|\lambda| = \Theta$ and $\lambda
= \Theta e^{2\pi i k /d}$.
\end{Satz}

The next proposition is a consequence of Perron-Frobenius and the first step
towards unique ergodicity of the system $(X(\zeta),T)$.

\begin{Satz} \label{PerronFrobConsequence}
Let $\zeta$ be a primitive substitution. For every $\alpha \in \kA$
the $s$-dimensional vector $\left (  \frac{L(\zeta^n(\alpha))}{\Theta^n} \right
)$ converges to the strictly positive eigenvector $v(\alpha)$ for the
PF-eigenvalue $\Theta$.
\end{Satz}
With that we get the following result.
\begin{Satz}
For every $\alpha \in \kA$ it holds that
$$
\frac{|\zeta^{n+1}(\alpha)|}{|\zeta^n(\alpha)|} \stackrel{n \rightarrow
\infty}{\longrightarrow} \Theta.
$$
\end{Satz}
\begin{Proof}
Using Proposition \ref{PerronFrobConsequence} it follows
$$
\frac{|\zeta^{n+1}(\alpha)|}{|\zeta^n(\alpha)|} = \frac{\langle
L(\zeta^{n+1}(\alpha)),\setId \rangle}{\Theta^{n+1}} \frac{\Theta^{n+1}}{\langle
L(\zeta^n(\alpha), \setId \rangle} \stackrel{n \rightarrow
\infty}{\longrightarrow} \frac{\langle v(\alpha), \setId \rangle}{\langle
v(\alpha), \setId \rangle} \Theta.
$$
\end{Proof}

The next proposition shows that every letter in $u$ appears with a positive
frequency in $u$ if $\zeta$ is a primitive substitution (in particular if
$\zeta$ fulfills the two conditions (\ref{FirstCond}) and (\ref{SecondCond})).

\begin{Satz} \label{LetterResult}
Let $\alpha,j \in \kA$, then it holds that
$$\frac{L_j (\zeta^n(\alpha))}{|\zeta^n(\alpha)|} \stackrel{n \rightarrow
\infty}{\longrightarrow} d_j,$$
where $d_j > 0$ is independent of $\alpha$.
\end{Satz}
\begin{Proof}
Using Proposition \ref{PerronFrobConsequence} and $|\zeta^n(\alpha)| = \langle
L(\zeta^n(\alpha)) , \setId \rangle$ we get
$$
\frac{L(\zeta^n(\alpha))}{|\zeta^n(\alpha)|} =
\frac{L(\zeta^n(\alpha))}{\Theta^n} \frac{\Theta^n}{\langle L(\zeta^n(\alpha)),
\setId \rangle} \stackrel{n \rightarrow \infty}{\rightarrow}
\frac{v(\alpha)}{\langle v(\alpha) , \setId \rangle}.
$$
The limit is the strictly positive and normed eigenvector of $\Theta$. Because
of that $d_j = v_j > 0$ and not depending on $\alpha$.
\end{Proof}

We now show that the system $(X(\zeta), T)$ is uniquely ergodic. A topological
dynamical system is called {\bf uniquely ergodic}, if there is an unique
$T$-invariant probability measure $\mu$ on $X$. In order to achieve that we
generalize the last result and replace primitivity with the more general
condition of minimality.

\begin{Satz}
Assume the system $(X,T)$ is minimal. Then for every letter $\alpha$ in the
fixed point $u$ and every word $B$ in $u$ it holds that
$$
\frac{L_B (\zeta^n(\alpha))}{|\zeta^n(\alpha)|} \stackrel{n \rightarrow
\infty}{\rightarrow} d_B,
$$
where $d_B > 0$ is not depending on $\alpha$.
\end{Satz}
\begin{Proof}
Let $B$ be a word in $u$ of length $l \ge 1$. For $l=1$ the claim follows
with Proposition \ref{LetterResult}. W.l.o.g. we assume $l \ge 2$ and define
$$
\Omega_l := \{B \in u: |B| = l \}.
$$
We show the claim while we use $\Omega_l$ as a new alphabet and define a
corresponding substitution $\zeta_l$ such that we can apply Proposition
\ref{LetterResult}. \\
Let $\omega$ be a letter in the new alphabet $\Omega_l$. We define
$\zeta_l : \Omega_l \rightarrow \Omega_l^*$ with the notation
$$
\zeta(\omega) = \zeta(\omega_0 \ldots \omega_{l-1}) = y_0 \ldots
y_{|\zeta(\omega_0)|-1} y_{|\zeta(\omega_0)|} \ldots y_{|\zeta(\omega)| -1},
$$
where $y_i \in \kA$ as
$$
\zeta_l(\omega) := (y_0 \ldots y_{l-1}) (y_1 \ldots y_l) \ldots (
y_{|\zeta(\omega_0)|-1} \ldots y_{|\zeta(\omega_0)|+l-2}).
$$
We can extend $\zeta_l$ via concatenation of symbols to $\Omega_l^*$
and $\Omega_l^\setN$. We now show the following two properties of $\zeta_l$.
\begin{itemize}
	\item [(a)] $\zeta_l$ has a fixed point $U_l \in \Omega_l^\setN$ with $U_l
= (u_0 \ldots u_{l-1}) (u_1 \ldots u_l) (u_2 \ldots u_{l+1}) \ldots$,
	\item [(b)] $\zeta_l$ is primitive, if $\zeta$ is primitive.
\end{itemize}
Proof of (a). \\
Let $\omega = u_0 \ldots u_{l-1}$ with $\zeta(\omega) = u_0 \ldots
u_{|\zeta(\omega)|-1}$ and $u = \zeta(u)$ we get
$$
\zeta_l(\omega) = (u_0 \ldots u_{l-1}) (u_1 \ldots u_l) \ldots (
u_{|\zeta(u_0)|-1} \ldots u_{|\zeta(u_0)| + l-2}).
$$
The word $\zeta_l(\omega)$ starts with $\omega$ and with
Proposition \ref{ExistenceFixpoint} the existence of a fixed point follows. For
every $n \ge 1$ we have
$$
\zeta_l^n(\omega) = (u_0 \ldots u_{l-1}) (u_1 \ldots u_l) \ldots
(u_{|\zeta^n(u_0)|-1} \ldots u_{|\zeta^n(u_0)|+l-2}),
$$
such that $\zeta_l^\infty(\omega) = U_l$.\\
Proof of (b). \\
With (a) $\zeta_l$ fulfills the conditions (\ref{FirstCond}) and
(\ref{SecondCond}) (use $u_0 \ldots u_{l-1}$ instead of $0$). Because of that
it is enough to show irreducibility of $\zeta_l$ on $\Omega_l$. Let
$\omega, B \in \Omega_l$. Since $u = \zeta^n(u)$ for every $n$, there is
an $\alpha \in \kA$ and $p \ge 1$ such that $B \subset \zeta^p(\alpha)$.
Because $\zeta$ is primitive it holds that $\alpha \in \zeta^m(\omega_0)$ for $m
\ge m_0$ and we get $B \subset \zeta^{m+p}(\omega_0)$. With the notation
$$
\zeta^n(\omega) = \zeta^n(\omega_0) \cdot \zeta^n(\omega_1, \ldots ,
\omega_{l-1}) = y_0 y_1 \ldots y_{|\zeta^n(\omega_0) |-1} \alpha_0 \alpha_1
\ldots,
$$
we obtain
\begin{equation} \label{zetaln}
\zeta_l^n(\omega) = (y_0 \ldots y_{l-1}) (y_1 \ldots y_l)\ldots (y_{|\zeta^n(\omega_0)|-1} \alpha_0 \ldots \alpha_{l-2}).
\end{equation}
$\zeta_l^n(\omega)$ contains all words of length $l$ which appear
in $\zeta^n(\omega_0)$. Choose $m$ big enough and define $n := m+p$,
then $\zeta_l^n(\omega)$ contains finally the word $B$ and the claim
is proven.

We now apply Proposition \ref{LetterResult} to $\zeta_l$ and obtain
$$
\lim_{n \rightarrow \infty} \frac{L_B(\zeta_l^n(\omega))}{|\zeta_l^n (\omega)|}
= d_B > 0,
$$
where $d_B$ is not depending on $\omega$. Obviously we have with (\ref{zetaln})
that $|\zeta_l^n(\omega)| = |\zeta^n(\omega_0)|$ and $L_B(\zeta_l^n(\omega))
\sim L_B(\zeta^n(\omega_0))$ for $n \rightarrow \infty$. Hence we get
$$
\lim_{n \rightarrow \infty} \frac{L_B (\zeta^n(\omega_0))}{|\zeta^n(\omega_0)|}
= d_B > 0.
$$
\end{Proof}

The value of $d_B$ is the frequency of the word $B$ in the fixed point $u$ and
is the $B$-entry of the normed eigenvector of the composition-matrix
for $\zeta_l$.

\begin{Bsp}
As an example we consider the Thue-Morse sequence with $M(\zeta) = M = \left
( \begin{array}{l l} 1 & 1 \\ 1 & 1 \end{array} \right )$ and eigenvalues
$\Theta = 2$, $\lambda = 0$. Define $\zeta_2$ on the alphabet $\Omega_2 =
\{(00),(01),(10),(11)\}$ like in the proof above
$$\zeta_2((00)) := (01) (10),$$
$$\zeta_2((01)) := (01) (11),$$
$$\zeta_2((10)) := (10) (00),$$
$$\zeta_2((11)) := (10) (01).$$
The composition-matrix $M_2$ for $\zeta_2$ is
$$
M_2 = \left ( \begin{array}{l l l l} 0 & 0 & 1 & 0 \\ 1 & 1 & 0 & 1 \\ 1 & 0 & 1
& 1 \\ 0 & 1 & 0 & 0 \end{array} \right ),
$$
with eigenvalues $\Theta = 2, \lambda = 0, 1, -1$. The normed eigenvector for
the eigenvalue $\Theta$ is $v = \left ( \frac{1}{6}, \frac{1}{3}, \frac{1}{3},
\frac{1}{6} \right )$, such that the frequencies of the pairs in $u$ are as
follows
$$
d_{(00)} = \frac{1}{6} =  d_{(11)}, \quad d_{(01)} = \frac{1}{3} = d_{(10)}.
$$
\end{Bsp}

Let $B$ be a word in $u$, then $[B]$ denotes the cylinderset which is
generated by $B$
$$
[B] := \{x \in X : x_j = b_j, \; 0 \le j \le |B| -1\}, \quad {\rm with} \; B =
b_0 \ldots b_{|B|-1}.
$$
Let $\mu$ be a $T$-invariant probability measure on $X$, then we can write $\mu$
as
$$
\mu([B]) = \lim_{j \rightarrow \infty} \frac{1}{N_j} |\{n < N_j: u_n \ldots
u_{n+|B|-1} = B \}|,
$$
for a sequence $(N_j)$ and every cylinderset $[B]$. In particular $\mu$ is for 
$N_j = |\zeta^j(\alpha)|$ a $T$-invariant probability measure and it holds that
$\mu([B]) = d_B$. The next proposition tell us that this measure is also
uniquely determined.

\begin{Satz}
If the system $(X,T)$ is minimal, then it is uniquely ergodic.
\end{Satz}

In particular we have the following 

\begin{Korollar}
Every vector $\mu = (\mu([j]))_{j=0}^{s-1}$ is the normed eigenvector to
the PF-eigenvalue $\Theta$.
\end{Korollar}

In the next step we want to investigate the composition-matrix $M_l$ of
$\zeta_l : \Omega_l \rightarrow \Omega_l$ and will derive an effective
method to calculate the frequencies of factors in $u$.
\begin{Satz}
Let $M= M(\zeta)$ be a primitive matrix with PF-eigenvalue $\Theta$, then $M_l$
is a primitive matrix with the same PF-eigenvalue for every $l \ge 2$.
\end{Satz}

We now show that we can derive the distribution of every word in $\Omega_l$
from the distribution of the words in $\Omega_2$. For that we fix $l \ge 2$.
Remark that for $p \ge 1$ it holds $(\zeta_l)^p = (\zeta^p)_l$, where
$(\zeta^p)_l: \Omega_l \rightarrow \Omega_l^*$ with $\omega = \omega_0 \ldots
\omega_{l-1} \in \Omega_l$ and
$$
\zeta^p(\omega) = \zeta^p(\omega_0) \ldots \zeta^p(\omega_{l-1}) = y_0 \ldots
y_{|\zeta^p(\omega)|-1},
$$
is defined as follows
$$
(\zeta^p)_l(\omega) := (y_0 \ldots y_{l-1}) (y_1 \ldots y_l) \ldots
(y_{|\zeta^p(\omega_0)|-1} \ldots y_{|\zeta^p(\omega_0)|+l-2}).
$$

If $p$ is greater than $l$, such that the condition
\begin{equation} \label{Boundedcondition}
|\zeta^p(\omega_0)| +l-2 < |\zeta^p(\omega_0)| + |\zeta^p(\omega_1)|
\end{equation}
is fulfilled, then $(\zeta^p)_l$ is completely determined through the
knowledge of the first two letters $\omega_0 \omega_1$ of $\omega$ on $\omega
\in \Omega_l$. Proposition \ref{PerronFrobConsequence} gives
$|\zeta^p(\omega_1) | \sim \Theta^p \Vert v(\omega_1) \Vert$ as $p
\rightarrow \infty$.
So we can express condition (\ref{Boundedcondition}) with
$$
\Theta^p > C \cdot l,
$$
where $C> 0$ is a constant. We now fix $p$ and $l$, such that condition
(\ref{Boundedcondition}) is fulfilled. Let $\pi_2 : \Omega_l \rightarrow
\Omega_2$ be the projection on the first two letters, that means $\pi_2(\omega_0
\ldots \omega_{l-1} ) = \omega_0 \omega_1$. We define $\tau_{2,l,p} : \Omega_2
\rightarrow \Omega_l^*$ with
$$
\tau_{2,l,p} (\omega_0 \omega_1) := (y_0 \ldots y_{l-1}) ( y_1 \ldots y_l)
\ldots (y_{|\zeta^p(\omega_0)|-1} \ldots y_{|\zeta^p(\omega_0)|+l-2}),
$$
if $\omega_0 \omega_1 \in \Omega_2$ and $\zeta^p(\omega_0 \omega_1) = y_0 \ldots
y_{|\zeta^p(\omega_0)|-1} y_{|\zeta^p(\omega_0)|} \ldots y_{|\zeta^p(\omega_0
\omega_1) |-1}$. Obviously it holds that
$$
\tau_{2,l,p} \circ \pi_2 = \zeta^p_l, \qquad \pi_2 \circ \tau_{2,l,p} =
\zeta^p_2,
$$
and
$$
\zeta_l \circ \tau_{2,l,p} = \tau_{2,l,p} \circ \zeta_2.
$$
We can extend the mappings $\tau_{2,l,p}$ and $\pi_2$ in a natural way to
mappings $\tau_{2,l,p}: \Omega_2^* \rightarrow \Omega_l^*$ and $\pi_2 :
\Omega_l^* \rightarrow \Omega_2^*$ and get the following commutative diagram

\xymatrixcolsep{7pc}
\xymatrixrowsep{5pc}
$$\xymatrix{\Omega_l^* \ar[d]^{\pi_2} \ar[r]^{\zeta_l^p} & \Omega_l^* \ar[r]^{\zeta_l} & \Omega_l^* \ar[d]^{\pi_2} \\
\Omega_2^* \ar[r]^{\zeta_2} \ar[ru]^{\tau_{2,l,p}} & \Omega_2^* \ar[r]^{\zeta_2^p} \ar[ru]^{\tau_{2,l,p}} & \Omega_2^*}$$

We write $L_l :\Omega_l^* \rightarrow \setR^{|\Omega_l|}$ for the
composition function of words $\omega$ in $\Omega_l$, then we get
$$
L_l(\zeta_l(\omega)) = M_l L_l(\omega), \qquad {\rm with} \; \omega \in
\Omega_l^*,
$$
and
$$
M_{2,l,p} L_2 (\pi_2 (\omega)) = L_l (\zeta_l^p(\omega)),
$$
where $M_{2,l,p}$ is the composition-matrix of $\tau_{2,l,p}$. With $A$ as a
matrix for the projection $\pi_2$ and $\xi = \pi_2(\omega)$ we get the
following commutative diagram

\xymatrixcolsep{7pc}
\xymatrixrowsep{5pc}
$$\xymatrix{L_l(\omega) \ar[d]^{A} \ar[r]^{M_l^p} & L_l(\zeta_l^p(\omega)) \ar[r]^{M_l} & L_l(\zeta_l^{p+1}(\omega)) \ar[d]^{A} \\
L_2(\xi) \ar[r]^{M_2} \ar[ru]^{M_{2,l,p}} & L_2(\zeta_2(\xi)) \ar[r]^{M_2^p} \ar[ru]^{M_{2,l,p}} & L_2(\zeta_2^{p+1}(\xi))}$$

\begin{Korollar}
The eigenvalues of $M_l$ coincide with the eigenvalues of $M_2$, if they are not
equal to zero.
\end{Korollar}
\begin{Proof}
Because of $M_{2,l,p} \cdot M_2 = M_l \cdot M_{2,l,p}$ it holds that for
every algebraic polynom
$$
M_{2,l,p} \cdot Q(M_2) = Q(M_l) \cdot M_{2,l,p}.
$$
On the other hand we have $M_{2,l,p} \cdot Q(M_2) A = Q(M_l)  \cdot M_l^p$,
such that for $Q(M_2) = 0$ the polynom $X \rightarrow Q(X) \cdot X^p$ leads
to vanishing of the matrix $M_l$. Furthermore $M_2^p Q(M_2) = A Q(M_l) M_l^p$
implies that $X \rightarrow Q(X) \cdot X^p$ leads to vanishing of the matrix
$M_2$, if $Q(M_l) = 0$.
\end{Proof}

\begin{Korollar}
If $v_2$ is an eigenvector of $M_2$ for the eigenvalue $\Theta$, then
$M_{2,l,p} \cdot v_2$ is an eigenvector of $M_l$ for the eigenvalue $\Theta$.
\end{Korollar}
\begin{Proof}
The claim follows from the fact that $M_{2,l,p} \cdot M_2 = M_l \cdot
M_{2,l,p}$.
\end{Proof}

For determining the frequency of a word $\omega \in \Omega_l$ of length $l$ it
is enough to determine the frequency for every pair $(\alpha \beta)$. Count how
often $\omega$ appear in $\zeta^p(\alpha \beta)$ under the condition that the
first letter of $\omega$ is in $\zeta^p(\alpha)$. This is then the entry
in $M_{2,l,p}$ on the position $(\omega, (\alpha \beta)) \in \Omega_l \times
\Omega_2$. If one consider for example the Thue-Morse sequence and want to
calculate the frequencies of words with length $5$ one has to set $p = 3$ so
that condition (\ref{Boundedcondition}) is fulfilled and get
$$\zeta^3(00) = 0110.1001.0110.1001,$$
$$\zeta^3(01) = 0110.1001.1001.0110,$$
$$\zeta^3(10) = 1001.0110.0110.1001,$$
$$\zeta^3(11) = 1001.0110.1001.0110.$$
There are $12$ words of length $5$ in $u$
$$(00101) \; (00110) \; (01001) \; (01011) \; (01100) \; (01101)$$
$$(11010) \; (11001) \; (10110) \; (10100) \; (10011) \; (10010).$$
The $M_{2,l,p}$-matrix has the following form
$$\left ( \begin{array}{l l l l} 1 & 0 & 1 & 1 \\ 0 & 1 & 1 & 0 \\ 1 & 1 & 0 & 1 \\ 1 & 0 & 1 & 1 \\ 0 & 1 & 1&  0 \\ 1 & 1 & 0 & 1 \\ 1&1&0&1 \\ 0&1&1&0 \\ 1&0&1&1\\ 1&1&0&1 \\ 0&1&1&0 \\ 1&0&1&1 \end{array} \right ).$$
Because of $v_2 = (1,2,2,1)$ we get $v_5 = M_{2,l,p} \cdot v_2 = (4,\ldots ,
4)$. Therefore every word has the frequency $\frac{1}{3 \cdot 2^2}$.
Analogous we can calculate the frequencies of words with arbitrary length.
\end{appendix}


\newpage


\begin{thebibliography}{6}
\addcontentsline{toc}{section}{Literatur}
	\bibitem[Bal10]{Bal10}
		{\sc Ball R. C.; Diakonova M.; MacKay R. S.}:
		{\em Quantifying Emergence in terms of Persistent Mutual Information},
		{Advances in Complex Systems, Vol. 13, No. 3, 327, (2010)}.
	\bibitem[Bec08]{Bec08}
		{\sc Beckermann Ansgar}:
		{\em Analytische Einführung in die Philosophie des Geistes},
		{Walter de Gruyter, 3. Auflage, (2008)}.
	\bibitem[Ber94]{Ber94}
		{\sc Berthé Valérie}:
		{\em Conditional entropy of some automatic sequences},
		{J. Phys. A, 27:7993-8006, (1994)}.
	\bibitem[Bil68]{Bil68}
		{\sc Billingsley Patrick}:
		{\em Convergence of Probability Measures},
		{John Wiley \& Sons, (1968)}.
	\bibitem[Bro25]{Bro25}
		{\sc Broad C.D.}:
		{\em The mind and its place in nature},
		{Kegan Paul, (1925)}.
	\bibitem[Cuc07]{Cuc07}
		{\sc Cucker Felipe; Smale, Steve}:
		{\em The Mathematics of Emergence},
		{Japanese Journal of Mathematics, Vol. 2, Nr. 1, (2007)}.
	\bibitem[Cha02]{Cha02}
		{\sc Chalmers D.J.}:
		{\em Strong and Weak Emergence},
		{Vol. The Re-Emergence of Emergence, Oxford University Press, (2002)}.
	\bibitem[Cru83]{Cru83}
		{\sc Crutchfield James; Packard N.H.}:
		{\em Symbolic dynamics of noisy chaos},
		{Physica D: Nonlinear Phenomena Vol. 7, Issue 1-3, 201 (1983)}.
	\bibitem[Cru94]{Cru94}
		{\sc Crutchfield James}:
		{\em The calculi of emergence: Computation, dynamics and induction},
		{Physica D, 75:11 54, (1994)}.
	\bibitem[Cru97]{Cru97}
		{\sc Crutchfield James; Feldman David}:
		{\em Statistical Complexity of Simple 1D Spin Systems},
		{Physical Review E 55:2, 1239R-1243R, (1997)}.
	\bibitem[Cru03]{Cru03}
		{\sc Crutchfield James; Feldman David}:
		{\em Regularities Unseen, Randomness Observed: Levels of Entropy Convergence},
		{Chaos, 15: 25-54. (2003)}.
	\bibitem[Cru10]{Cru10}
		{\sc Crutchfield James; Ellison Christopher; James Ryan; Mahoney John}:
		{\em Synchronization and Control in Intrinsic and Designed Computation: An Information-Theoretic Analysis of Competing Models of 
		Stochastic Computation},
		{Santa Fe Institute Working Paper 10-07-XXX. (2010)}.
	\bibitem[Cov06]{Cov06}
		{\sc Cover Thomas; Thomas Joy}:
		{\em Elements of Information theory},
		{John Wiley \& Sons, Second Edition, (2006)}.
	\bibitem[Dek92]{Dek92}
		{\sc Dekking F M}:
		{\em On the Prouhet-Thue-Morse Measure},
		{Acta Universitatis Carolinae, Mathematica et Physica 33 35-40, (1992)}.
	\bibitem[deL89]{deL89}
		{\sc de Luca A; Varrichio S}:
		{\em Some combinatorical properties of the Thue-Morse sequence},
		{Theor. Comput. Sci. 63, 333-348, (1989)}.
	\bibitem[Ell09]{Ell09}
		{\sc Ellison Christopher; Mahoney John; Crutchfield James}:
		{\em Prediction, Retrodiction and the amount of Information stored in the Present},
		{Journal of Statistical Physics, Vol. 136, Nr. 6, (2009)}.
	\bibitem[Fel98]{Fel98}
		{\sc Feldman David; Crutchfield James}:
		{\em Discovering noncritical organization: Statistical mechanical, information theoretic and computational views of patterns in simple one-		dimensional spin systems},
		{Santa Fe Institute Working Paper 98-04-026 (1998)}.
	\bibitem[Fog08]{Fog08}
		{\sc Fogg N. Pytheas}:
		{\em Substitutions in Dynamics, Arithmetics and Combinatorics},
		{Springer-Verlag (2008)}.
	\bibitem[Gme10]{Gme10}
		{\sc Gmeiner Peter}:
		{\em Komplexitätsmaße und Emergenz},
		{Diploma-Thesis (in German), Erlangen, (2010)}.
	\bibitem[Gra86]{Gra86}
		{\sc Grassberger Peter}:
		{\em Toward a quantitative theory of self-generated complexity},
		{International Journal of Theoretical Physics, Volume 25, Issue 9, pp.907-938, (1986)}.
	\bibitem[Gra90]{Gra90}
		{\sc Gray Robert}:
		{\em Entropy and Information Theory},
		{Springer-Verlag, (1990)}.
	\bibitem[Guc83]{Guc83}
		{\sc Guckenheimer J.; Holmes P.}:
		{\em Nonlinear Oscillations, Dynamical Systems and Bifurcation of Vector
Fields},
		{Springer-Verlag, (1983)}.
	\bibitem[Jet89]{Jet89}
		{\sc Jetschke, Gottfried}:
		{\em Mathematik der Selbstorganisation},
		{Deutscher Verlag der Wissenschaften, (1989)}.
	\bibitem[Hak83]{Hak83}
		{\sc Haken Hermann}:
		{\em Synergetik. Eine Einführung},
		{Springer-Verlag, 2. Auflage, (1983)}.
	\bibitem[Kna06]{Kna06}
		{\sc Knauf Andreas; Seiler Ruedi}:
		{\em Vorlesungsskript zur Statistischen Mechanik},
		{Wintersemester 2006/07}.
	\bibitem[Loe10]{Loe10}
		{\sc L\"ohr Wolfgang}:
		{\em Models of Discrete-Time Stochastic Processes and Associated
Complexity Measures},
		{PhD-Thesis, Leipzig, (2010)}.
	\bibitem[Lon00]{Lon00}
		{\sc London Franz}:
		{\em Ueber Doppelfolgen und Doppelreihen},
		{Mathematische Annalen, Vol. 53, Nr. 3,  322-370, Springer, (1900)}.
	\bibitem[Pin64]{Pin64}
		{\sc Pinsker M.S.}:
		{\em Information and Information Stability of Random Variables and Processes},
		{Holden-Day, Inc., (1964)}.
	\bibitem[Pol04]{Pol04}
		{\sc Polani Daniel}:
		{\em Defining Emergent Descriptions by Information Preservation},
		{InterJournal, Complex Systems 1102, (2004)}.
	\bibitem[Pol06]{Pol06}
		{\sc Polani Daniel}:
		{\em Emergence, Intrinsic Structure of Information and Agenthood},
		{InterJournal, Complex Systems 1937, (2006)}.
	\bibitem[Que87]{Que87}
		{\sc Queffélec Martine}:
		{\em Substitution Dynamical Systems - Spectral Analysis},
		{Springer-Verlag, (1987)}.
	\bibitem[Set08]{Set08}
		{\sc Seth, Anil K.}:
		{\em Measuring emergence via nonlinear Granger causality},
		{In: Bullock S, Watson R, Noble J, Bedau M, editors. Artificial life XI:
proceedings of the 11th international conference on the simulation and 	
synthesis of living systems. Cambridge MIT Press, 545-552, (2008)}.
	\bibitem[Sha01]{Sha01}
		{\sc Shalizi Cosma Rohilla}:
		{\em Causal Architecture, Complexity and Self-Organization in Time},
		{PhD-Thesis, (2001)}.
	\bibitem[Sta91]{Sta91}
		{\sc Stanley, Jeanette; Bak Evan}:
		{\em Neuronale Netze},
		{Systhema Verlag, (1991)}.
	\bibitem[Ste99]{Ste99}
		{\sc Stephan, Achim}:
		{\em Emergenz. Von der Unvorhersagbarkeit zur Selbstorganisation},
		{Dresden University Press, 1. Auflage, (1999)}.
	\bibitem[Yeu91]{Yeu91}
		{\sc Yeung Raymond}:
		{\em A New Outlook on Shannon's Information Measures},
		{IEEE Transactions on Information Theory, Vol. 37, No. 3, (1991)}.
\end{thebibliography}
\end{document}